\documentclass[aps,
prX,
twocolumn,
longbibliography,
amsmath,
amssymb]{revtex4-2}
 
\usepackage{graphicx}
\usepackage[normalem]{ulem}
\usepackage{soul}
\usepackage{dcolumn}
\usepackage{bm}
\usepackage{multirow}
\usepackage{esint}
\usepackage[utf8]{inputenc}
\usepackage{relsize}
\usepackage{color}
\usepackage{siunitx}
\usepackage[thinlines]{easytable}
\usepackage{cleveref}
\usepackage{amssymb}

\newcommand{\red}{\color{red}}

\newcommand{\blue}{\color{blue}}

\usepackage{siunitx}
\usepackage{relsize}
\usepackage{blindtext}
\usepackage[thinlines]{easytable}
\usepackage{bm}
\usepackage{lipsum}
\usepackage[section]{placeins}

\begin{document}

\preprint{APS/123-QED}

\title{Theory of magnetic spin and orbital Hall and Nernst effects in bulk ferromagnets}

\author{Leandro Salemi}
 \email{leandro.salemi@physics.uu.se}
\author{Peter M.\ Oppeneer}%
\affiliation{%
 Department of Physics and Astronomy, Uppsala University, P.\ O.\ Box 516, SE-75120 Uppsala, Sweden
 }%

\date{\today}

\begin{abstract}
The magnetic spin Hall effect (MSHE) is an anomalous charge-to-spin conversion phenomenon which occurs in ferromagnetic materials. In contrast to the conventional spin Hall effect (SHE), being a {time-reversal even} effect, the magnetic counterpart is time-reversal odd.

In this work, we use \textit{ab initio} calculations to investigate the MSHE for the bulk ferromagnets Fe, Co, and Ni. The magnitudes of the MSHE of Fe and Co are comparable to those of the SHE, but the MSHE is strongly dependent on the electron lifetime and the MSHE and SHE can moreover have opposite signs. For Ni the MSHE is smaller than the SHE, but in general, the MSHE cannot be ignored for spin-orbit torques.  Considering a charge current we analyze how both the MSHE and SHE contribute to a total Hall angle. We extend our analysis of the MSHE to its orbital counterpart, that is, the magnetic orbital Hall effect (MOHE), for which we show that the MOHE is in general smaller than the orbital Hall effect (OHE). We compute furthermore the thermal analogs, i.e., the spin and orbital Nernst effects, and their magnetic counterparts. Here our calculations show that the magnetic spin and orbital Nernst effects of Ni are substantially larger than those of Fe and Co.
\end{abstract}

\pacs{Valid PACS appear here}
\maketitle

\section{Introduction}
Understanding the generation of spin currents at the microscopic scale is a fundamental issue in the field of spintronics. The spin Hall effect (SHE) is one of the most promising phenomena in this field that has 
captivated the scientific community since the early 2000s \cite{Sinova2015,Hoffmann2013}. In its conventional definition, the SHE describes the electrical generation of a spin current, where the electric field $\bm{E}$, spin current $\bm{J}^{\bm{S}}$, and induced spin polarization $\bm{S}$ are mutually orthogonal.

The SHE was theoretically
proposed half a century ago by Dyakonov and Perel \cite{dyakonovPossibilityOrientingElectron1971, dyakonovCurrentinducedSpinOrientation1971}, but did not attract much attention until a 1999 letter by Hirsch, whose title would give its name to this effect \cite{hirschSpinHallEffect1999}. There, Hirsch predicted that spin-orbit scattering centers would give rise to an electrically-generated transverse spin current which would lead to spin accumulation at the edges of nonmagnetic metals. Soon after, it was shown that spin diffusion using a semiclassical Boltzmann approach would also lead to spin accumulation \cite{zhangSpinHallEffect2000}.

Experimentally, the SHE was first observed in semiconducting materials \cite{katoCoherentSpinManipulation2004,katoCurrentInducedSpinPolarization2004,stephensSpinAccumulationForwardBiased2004}. Effects orders of magnitude larger were later observed in heavy-metals like Pt, via the SHE as well as its inverse effect, the inverse SHE (ISHE) \cite{saitohConversionSpinCurrent2006,kimuraRoomTemperatureReversibleSpin2007, sekiGiantSpinHall2008,Stamm2017}. The impressive interest in SHE-related phenomena is strongly rooted in its practicality, as it has been experimentally proven over the last decade that SHE-generated spin currents could be used to reversibly and efficiently control magnetization \cite{MihaiMiron2011,liuCurrentInducedSwitchingPerpendicularly2012, jamaliSpinOrbitTorquesCo2013, fanMagnetizationSwitchingGiant2014, garelloUltrafastMagnetizationSwitching2014, haoGiantSpinHall2015, leeThermallyActivatedSwitching2014}.

The microscopic origin of the SHE can be decomposed into an intrinsic and extrinsic contribution. The intrinsic contributions originates from the spin Berry curvature associated to the band structure topology of the material \cite{karplusHallEffectFerromagnetics1954, murakamiDissipationlessQuantumSpin2003, sinovaUniversalIntrinsicSpin2004, tanakaIntrinsicSpinHall2008}, while the extrinsic mechanisms, such as skew-scattering and side jumps, emerge from spin-dependent scattering on defects, as proposed for the anomalous Hall effect \cite{bergerSideJumpMechanismHall1970, smitSpontaneousHallEffect1958}.

The concept of the SHE can be extended to orbital angular momentum, leading to the orbital Hall effect (OHE). While the observation of orbital transport is a topic of on-going efforts, theoretical investigations have shown that a huge intrinsic OHE arises in Pt, without requiring spin-orbit coupling (SOC) \cite{kontaniStudyIntrinsicSpin2007}. Other theoretical investigations of the OHE were later conducted and similar observations were made \cite{tanakaIntrinsicSpinHall2008, kontaniGiantOrbitalHall2009, kontaniGiantIntrinsicSpin2008, tanakaIntrinsicSpinOrbital2010, goIntrinsicSpinOrbital2018, joGiganticIntrinsicOrbital2018}. Similarly to the SHE, it is often assumed that the electric field $\bm{E}$, orbital current $\bm{J}^{\bm{L}}$ and orbital polarization $\bm{L}$ are mutually orthogonal.

Another variant of the SHE has emerged in recent years. The $3^{\rm rd}$-rank spin Hall tensor $\sigma_{ij}^{S_k}$, with Cartesian indices $i$, $j$, and $k$, is uniquely defined for nonmagnetic metals with cubic crystal symmetry,
as 
\begin{equation}
\label{eq:spin-hall-conductivity-paramagnetic-cubic}
\sigma^{S_k}_{ij} = \epsilon_{ijk} \, \sigma_{_\text{SH}},
\end{equation}
with $\epsilon_{ijk}$ the Levi-Civita tensor. The SHE is then described by a single isotropic quantity, the spin Hall conductivity (SHC) $\sigma_{_{\rm SH}}$ which is time-reversal invariant.
However, it has become evident in the last years that the SHE is not only determined by the crystal structure, but also by the appearance of magnetic order. The latter not only can break spatial symmetry (e.g., ferromagnetism) but also breaks time-reversal symmetry, which can give rise to the appearance of nonzero, $\mathcal{T}$-odd components in $\bm{\sigma}^{\bm{S}}$. 

Signatures of such unusual $\mathcal{T}$-odd components have been recently observed \cite{humphriesObservationSpinorbitEffects2017, Wang2019, davidsonPerspectivesElectricallyGenerated2020}. They were recently discussed in the case of non-collinear antiferromagnets \cite{zeleznySpinPolarizedCurrentNoncollinear2017, zeleznySpinTransportSpin2018, kimataMagneticMagneticInverse2019}. Such $\mathcal{T}$-odd generation of spin currents has been referred to as magnetic SHE (MSHE)  \cite{kimataMagneticMagneticInverse2019,  mookOriginMagneticSpin2020, Salemi2021}. Although it was proposed that such components should exists for a broader class of materials, such as simple ferromagnets \cite{seemannSymmetryimposedShapeLinear2015,Wang2021}, no material-dependent \textit{ab initio} study has been performed for these so far.

In this paper, we use relativistic electronic structure calculations within the linear-response framework to investigate the magnetic spin and orbital conductivities for bcc Fe, hcp Co, and fcc Ni. We compute the full anisotropic $\bm{\sigma}^{\bm{S}}$ tensor and quantify the SHE and MSHE components. We also predict the orbital analog to the MSHE, that is, the magnetic OHE (MOHE), which has not yet been observed. We compute the full $\bm{\sigma}^{\bm{L}}$ tensor which allows us to fully quantify the MOHE, as well as its anisotropy. We then extend our discussion to consider thermally-driven spin and orbital current generation and compute the magnetic counter part of the recently observed spin Nernst effect (SNE) \cite{Meyer2017,Sheng2017,Bose2018}, i.e., the magnetic SNE (MSNE), as well as an orbital Nernst effect (ONE) and magnetic orbital Nernst effect (MONE). Our calculations show that the MSHE, in particular, is comparable in size to the SHE, but can have opposite sign. It needs therefore to be taken into account when electrically-induced spin currents in ferromagnetic materials or heterostructures are investigated.

In the following we first introduce the theoretical framework in Sec.\ \ref{Theory}, followed by the presentation of calculated results in Sec.\ \ref{Results}. Implications of the results are discussed in Sec.\ \ref{Discussion}.

\section{Theory}
\label{Theory}
\subsection{Symmetry considerations}

The electrical generation of spin currents is quantified by the 3$^{\text{rd}}$ rank spin conductivity tensor $\bm{\sigma}^{\bm{S}}$, which relates the 2$^{\text{nd}}$ rank spin current density tensor $\bm{J}^{\bm{S}}$ to the external electric field $\bm{E}$,
\begin{equation}
\label{eq:spincurrent-electricfield-relation}
J_i^{S_k} = \sigma^{S_k}_{ij} E_j,
\end{equation}
for the Cartesian indices $i$, $j$, and $k$. 

Note that we focus here on the spin angular momentum, but without loss of generality a similar formulation can be straightforwardly extended to the orbital angular momentum.
Conventionally, the SHE relates to the time-reversal even ($\mathcal{T}$-even) anti-symmetric part of $\bm{\sigma}^{\bm{S}}$. Because there exists no crystal symmetry for which all components of $\bm{\sigma}^{\bm{S}}$ vanish, the SHE can always be observed in any material. In nonmagnetic cubic materials the high symmetry of the crystal structure imposes that only one quantity, the spin Hall conductivity 
$\sigma_{_{\rm{SH}}}$ remains, see Eq.\ (\ref{eq:spin-hall-conductivity-paramagnetic-cubic}).

In the presence of magnetism the situation is different, due to the lowering of symmetry by the magnetization. The symmetry of the SHE tensor has been analyzed previously for different crystal symmetries
\cite{seemannSymmetryimposedShapeLinear2015,Zelezny2017}. As our aim is here to study ferromagnetic bcc Fe, hcp Co, and fcc Ni we consider the specific nonzero tensor elements of $\bm{\sigma}^{\bm{S}}$ for these materials. In addition, we choose the magnetic moment $\bm{M}$ along the (001) crystallographic direction for Fe and Ni ($4/mm'm'$ magnetic Laue group) and the (0001) direction for hcp Co ($6/mm'm'$). This direction we define as the $\bm{u}_z$ direction. 
The tensor $\bm{\sigma}^{\bm{S}}$ can then be written as
\begin{subequations}
\begin{align}
\label{eq:sigma-Sx-cubic-explicit}
\bm{\sigma}^{S_x} &= 
\begin{pmatrix}
0 & 0 & {\red \sigma^{S_x}_{xz}} \\
0 & 0 & {\blue \sigma^{S_x}_{yz}} \\
{\red \sigma^{S_x}_{zx}} & {\blue \sigma^{S_x}_{zy}} & 0 \\
\end{pmatrix},\\
\label{eq:sigma-Sy-cubic-explicit}
\bm{\sigma}^{S_y} &= 
\begin{pmatrix}
0 & 0 & {\blue \sigma^{S_y}_{xz}} \\
0 & 0 & {\red \sigma^{S_y}_{yz}} \\
{\blue \sigma^{S_y}_{zx}} & {\red \sigma^{S_y}_{zy}} & 0 \\
\end{pmatrix},\\
\label{eq:sigma-Sz-cubic-explicit}
\bm{\sigma}^{S_z} &= 
\begin{pmatrix}
\sigma^{S_z}_{xx} & {\blue \sigma^{S_z}_{xy}} & 0 \\
{\blue \sigma^{S_z}_{yx}} & \sigma^{S_z}_{yy} & 0 \\
0 & 0 & \sigma^{S_z}_{zz}\\
\end{pmatrix}.
\end{align}
\end{subequations}

The components in the tensor can be divided in three categories. 
First, the components $\sigma^{S_k}_{ij}$ where $\epsilon_{ijk} \neq 0$ can be referred to as SHE-like because (1) $\bm{J}$, $\bm{J}^{\bm{S}}$, and the spin polarization direction of $\bm{J}^{\bm{S}}$ are mutually orthogonal, and (2) they are even upon time-reversal symmetry ($\mathcal{T}$-even). These elements are indicated with blue color in Eqs.\  (\ref{eq:sigma-Sx-cubic-explicit})-(\ref{eq:sigma-Sz-cubic-explicit}). Contrarily to the case of nonmagnetic cubic materials, we have $\sigma^{S_z}_{xy} \neq \sigma^{S_y}_{zx} \neq \sigma^{S_x}_{yz}$, due to the magnetism-induced lowering of symmetry. As a consequence, there is not a single SHC as the relative orientation of $\bm{M}$ and the spin polarization direction of $\bm{J}^{\bm{S}}$ enters the picture. There are nevertheless further symmetry relations: $\sigma_{zx}^{S_y} = - \sigma_{zy}^{S_x}$, $\sigma_{yz}^{S_x}= -\sigma_{xz}^{S_y}$, and $\sigma_{xy}^{S_z} = - \sigma_{yx}^{S_z}$. 

Second, the components $\sigma^{S_x}_{xz}$, $\sigma^{S_x}_{zx}$, $\sigma^{S_y}_{yz}$ and $\sigma^{S_y}_{zy}$, shown with red color that can be referred to as MSHE-like. These emerge from the ferromagnetism-induced lowering of symmetry, are odd upon time-reversal symmetry ($\mathcal{T}$-odd) and require spin-orbit coupling (SOC) to exist. Signature of MSHE components have been observed in recent experimental works \cite{humphriesObservationSpinorbitEffects2017, Wang2019, kimataMagneticMagneticInverse2019}. However,  only few materials' dependent \emph{ab initio} calculations (e.g., \cite{mookOriginMagneticSpin2020}) have investigated them so far. One of the main result of this work is the estimation of those anomalous components.

Finally, we have the diagonal components of $\bm{\sigma}^{S_z}$, that is, $\sigma^{S_z}_{xx}$, $\sigma^{S_z}_{yy}$, and $\sigma^{S_z}_{zz}$ [black diagonal elements in Eqs.\  (\ref{eq:sigma-Sx-cubic-explicit})-(\ref{eq:sigma-Sz-cubic-explicit})]. Although they are $\mathcal{T}$-odd like the MSHE components, their physical origin is very different. They emerge from the difference in the longitudinal conductivity of spin-up and spin-down electrons, and would still exist if SOC is turned off. They lead to a spin-polarized conductivity, similar to the spin-dependent Seebeck effect \cite{Dejene2012} that quantifies the charge transport driven by a thermal gradient in a ferromagnet.

The unusual MSHE components $\sigma_{zx}^{S_x}$ and $\sigma_{zy}^{S_y}$ induce, for an $x$-$y$-plane electric field $\bm{E}$, a spin current $\bm{J}^{\bm{S}}$ parallel to $\bm{M}$, but the spin polarization is directed along $\bm{E}$. The  $\mathcal{T}$-even elements also lead to a spin current $\bm{J}^{\bm{S}} \, || \, \bm{M}$, but with spin polarization perpendicular to $\bm{E}$. The two induced spin polarizations will thus exert torques in orthonormal directions.

\subsection{Computational methodology}

To compute the spin and orbital Hall tensors, we use relativistic density-functional theory (DFT) as implemented in the all-electron, full-potential code WIEN2k \cite{Blaha2018}. The calculated Kohn-Sham eigenstates $| n \bm{k} \rangle$ and band eigenenergies $\varepsilon_{n\bm{k}}$, with $n$ the band index and $\bm{k}$ the wavevector, are used as input for the linear-response theory calculations. The Kubo linear-response expression \cite{Kubo1957,Salemi2021} for the spin Hall tensor reads
\begin{equation}
\label{eq:LinearResponse}
\sigma_{ij}^{S_k} = -\frac{ie \hbar}{m_e} \int_{\Omega} \frac{d\bm{k}}{\Omega}
\sum_{n, m} \frac{f_{m\bm{k}} - f_{n\bm{k}} }{\varepsilon_{m\bm{k}} -\varepsilon_{n\bm{k}}}~
\frac{J_{i,mn\bm{k}}^{S_k} ~ p_{j,nm\bm{k}} }{\varepsilon_{m\bm{k}}-\varepsilon_{n\bm{k}} + i\delta} ,
\end{equation}
where $f_{n\bm{k}}$ is the Fermi-Dirac function,  $m_e$ the electron mass, $\Omega$ the Brillouin zone volume, and $p_{j,nm\mathbf{k}}$ the $j^\text{th}$ component of the momentum-operator ($\hat{\bm{p}}$) matrix element.
The quantity in the sum over band indices is called the spin Berry curvature (for $n \neq m$).
$ \hat{J}^{\hat{S}_k}_{i,mn\bm{k}}$ is the matrix element of the spin current operator, given by 
\begin{equation}
{J}^{\hat{S}_k}_{i} = \frac{\{\hat{S}_k, \hat{p}_i\}}{2Vm_e},
\end{equation}
with $\hat{S}_k$ the spin operator, $V$ the volume of the unit cell,  and $\{\hat{A},\hat{B}\} = \hat{A}\hat{B} + \hat{B}\hat{A}$ is the anti-commutator.
 The parameter $\delta= \tau^{-1}$ describes the finite electron lifetime due to scattering events. It can in principle depend on the band indices and whether the transition in the $n,m$ sum is from an intraband $n=m $ or an interband ($n \neq m$) transition. We will initially set $\delta$ to $40$ meV and show the lifetime broadening dependence in Sec.\  \ref{Results}. The $\bm{k}$ sums in Eq.\ (\ref{eq:LinearResponse}) are numerically evaluated on $k$-meshes containing at least $2\,10^{4}$ $k$ points.

The same formalism can be directly used to compute the orbital Hall conductivity tensor $\bm{\sigma}^{\bm{L}}$ for which one has to replace $\hat{S}_k$ by the orbital angular momentum, $\hat{L}_k$. The SHC and OHC tensors can in addition be evaluated as a function of the band filling, by varying the electrochemical potential $E$  which is contained in the occupation function $f_{n\bm{k}}$. 

Once $\boldsymbol{\sigma}^{S_k}(E)$ and $\boldsymbol{\sigma}^{L_k}(E)$ have been computed, magnetothermal transport coefficients $\Lambda^{S_k(L_k)}_{ij}$ can then be derived from these using the Mott formula \cite{cutlerObservationAndersonLocalization1969},
\begin{equation}
\label{eq:Mott_Formula}
\Lambda^{S_k(L_k)}_{ij} = \frac{\pi^2 k_B^2 T}{-3e} \Big( \frac{d}{dE} \sigma^{S_k(L_k)}_{ij} (E) \Big)_{E=E_F},
\end{equation}
where $k_B$ is the Boltzmann constant, $T$ the temperature in Kelvin and $e > 0$ the elementary charge. The derivative is taken with respect to the electrochemical potential $E$ in $f_{n\bm{k}}$. By definition, $E=0$ corresponds to the Fermi level.

\section{Results}
\label{Results}
\subsection{Spin and orbital Hall effect ($\mathcal{T}$-even)}
Let us first focus on the SHE and OHE, which we respectively define as the electrical generation of a spin and orbital current arising from the $\mathcal{T}$-even components of $\bm{\sigma}^{\bm{S}}$ and $\bm{\sigma}^{\bm{L}}$. For the considered materials, those are the $\sigma^{S_k(L_k)}_{ij}$ tensor components such that $\epsilon_{ijk} \neq 0$.

We set the magnetization $\bm{M}$ along the (001) crystallographic direction for Fe and Ni, and along (0001) for hcp Co, and choose this to be the $\bm{u}_z$ direction. This leads to three components that are not invariant under cyclic permutation. Specifically, these are 
\begin{itemize}
\itemsep0em 
\item[-] $ \sigma^{S_y(L_y)}_{zx} =  - \sigma^{S_x(L_x)}_{zy} $: components where the flow of spin (orbital) current $\bm{J}^{\bm{S}}$ ($\bm{J}^{\bm{L}}$) is parallel to $\bm{M}$.
\item[-] $  \sigma^{S_x(L_x)}_{yz} = - \sigma^{S_y(L_y)}_{xz} $: components where the externally applied electric field $\bm{E}$ is parallel to $\bm{M}$.
\item[-] $  \sigma^{S_z(L_z)}_{xy} = - \sigma^{S_z(L_z)}_{yx} $: components where the spin (orbital) polarization $\bm{S}$ ($\bm{L}$) is parallel to $\bm{M}$.
\end{itemize}

\begin{figure}[tbt!]
  \includegraphics[width=\linewidth]{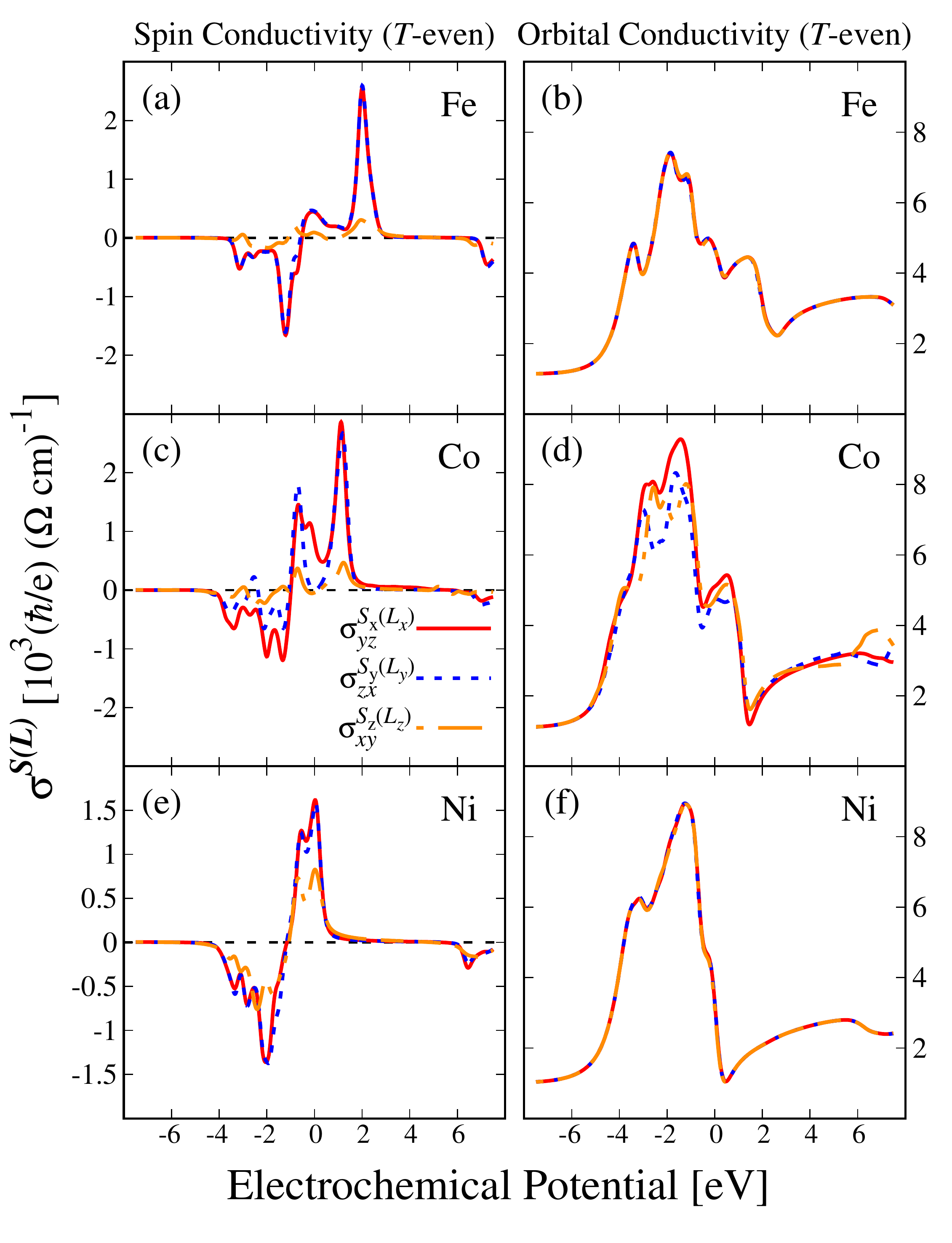}
  \caption{\textit{Ab initio} calculated $\mathcal{T}$-even components of $\bm{\sigma}^{\bm{S}}$ as function of the electrochemical potential for (a) Fe, (c) Co, and (e) Ni, and of $\bm{\sigma}^{\bm{L}}$ for (b) Fe, (d) Co, and (f) Ni. The considered components have indices $i$, $j$, and $k$ such that $\epsilon_{ijk} \neq 0$. The lifetime broadening used is $\hbar \tau^{-1} = 40$ meV.}
  \label{fig:SHE-like-Conductivities}
\end{figure}

In  Fig.\ \ref{fig:SHE-like-Conductivities} we show the calculated results for those components for ferromagnetic Fe, Co, and Ni, as a function of the electrochemical potential $E$. Focusing first on the spin conductivity (left-hand column in Fig.\ \ref{fig:SHE-like-Conductivities}), we clearly notice that the components for which $\bm{S}$ is orthogonal to $\bm{M}$, that is $\sigma^{S_y}_{zx(xz)}$ and $\sigma^{S_x}_{yz(zy)}$, show higher absolute-value maxima than $\sigma^{S_z}_{xy(yx)}$. For the rightmost peak, $\sigma^{S_y}_{zx}$ is two to eight times larger than  $\sigma^{S_z}_{xy}$. This emphasizes that the $\bm{M}$-induced lowering of symmetry cannot be neglected for the SHE, even for simple ferromagnets.

For Fe and Ni, the components $\sigma^{S_y}_{zx}$ and $\sigma^{S_x}_{yz}$ are nearly identical, though not equal. In this case, the SHE-like spin conductivity can be, in a good approximation, split into two components, depending whether the spin polarization of the spin current is parallel ($\sigma^{S_z}_{xy(yx)}$) or perpendicular ($\sigma^{S_x}_{yz(zy)}$ and $\sigma^{S_y}_{zx(xz)}$) to $\bm{M}$. For hcp Co, all components significantly differ from each other, suggesting that structural asymmetry has a greater impact than the magnetic asymmetry.

The OHE-like components (right-hand column in Fig.~\ref{fig:SHE-like-Conductivities}) are, in a peak-to-peak comparison, several times to one order of magnitude larger than their SHE-like analogs. Contrarily to the spin components, for Fe and Ni no substantial difference can be observed between $\sigma^{L_x}_{yz}$, $\sigma^{L_y}_{zx}$, and $\sigma^{L_z}_{xy}$. Those components are however noticeably different for Co, stressing that the structural asymmetry influences the OHE significantly, whereas the $\bm{M}$-induced asymmetry has virtually no effect on the OHE. It deserves to be mentioned once more that the OHE components are present even when the SOC is set to zero \cite{tanakaIntrinsicSpinHall2008, goIntrinsicSpinOrbital2018}, whereas SHE-like components vanish. 

The calculated values for the SHE and OHE components at the Fermi level are given in Table~\ref{tab:SHE-OHE-Conductivities}, in units of $\frac{\hbar}{e} (\Omega \,\textrm{cm})^{-1}$.  While Ni shows the smallest $\sigma^{S_x}_{yz} / \sigma^{S_z}_{xy}$ ratio,  the absolute value of $\sigma^{S_x}_{yz}$ is remarkably high. For instance, one could compare to $\sigma^{S_z}_{xy} \approx 2000 \,\frac{\hbar}{e} (\Omega \, \text{cm})^{-1}$ calculated for Pt \cite{Guo2008,Stamm2017}, which is often considered as a material of choice when it comes to SHE-based generation of spin currents. The anisotropy of the three SHE components of hcp Co is predicted to be huge. It should be possible to observe such anisotropy in SHE measurements on single-crystalline Co. The OHE-like components at the Fermi energy are in contrast quite isotropic and substantially larger than the SHE-like components.

\begin{table}[thb!]
\begin{ruledtabular}
\caption{\textit{Ab initio} calculated values for the SHE- and OHE-like components of the spin and orbital conductivity tensors, as well as their magnetic components MSHE and MOHE, for ferromagnetic bcc Fe, hcp Co, and fcc Ni, in units of $\frac{\hbar}{e} (\Omega \,\text{cm})^{-1}$. The magnetization is chosen along the $z$ axis and the lifetime broadening $\hbar \tau^{-1} = 40$ meV. }
\label{tab:SHE-OHE-Conductivities}
\centering
\begin{tabular}{lccc c ccc c cc c cc}
      & \multicolumn{3}{c}{SHE} & & \multicolumn{3}{c}{OHE} & & \multicolumn{2}{c}{MSHE} && \multicolumn{2}{c}{MOHE} \\
      \cline{2-4} \cline{6-8} \cline{10-11} \cline{13-14} \tabularnewline[-3\doublerulesep] 
      &  $\sigma^{S_x}_{yz}$  &  $\sigma^{S_y}_{zx}$  &   $\sigma^{S_z}_{xy}$ & &  $\sigma^{L_x}_{yz}$  &  $\sigma^{L_y}_{zx}$  &   $\sigma^{L_z}_{xy}$  & & $\sigma_{xz}^{S_x}$  & $\sigma_{zx}^{S_x}$  & & $\sigma_{xz}^{L_x}$  & $\sigma_{zx}^{L_x}$
      \tabularnewline[\doublerulesep]
      \hline     
      \tabularnewline[-3\doublerulesep] 
  Fe  &  441  &  456  &   92 &  & 4697  &  4698  &   4707  & & -593 & 739 & & 1343 & 848 \\[0.35em]
  Co  &  839  &  8 &  -44 &  & 5103  &  4718  &  4737   & & 614 & 1074 & & -358 & 1356  \\[0.35em]
  Ni  &  1606  &  1543  &  824 &  & 3306  &  3297  &  3149   & & 394 & -290 & & -66 & 1033 \\
\end{tabular}
\end{ruledtabular}
\end{table}

\begin{figure}[b!]
  \includegraphics[width=\linewidth]{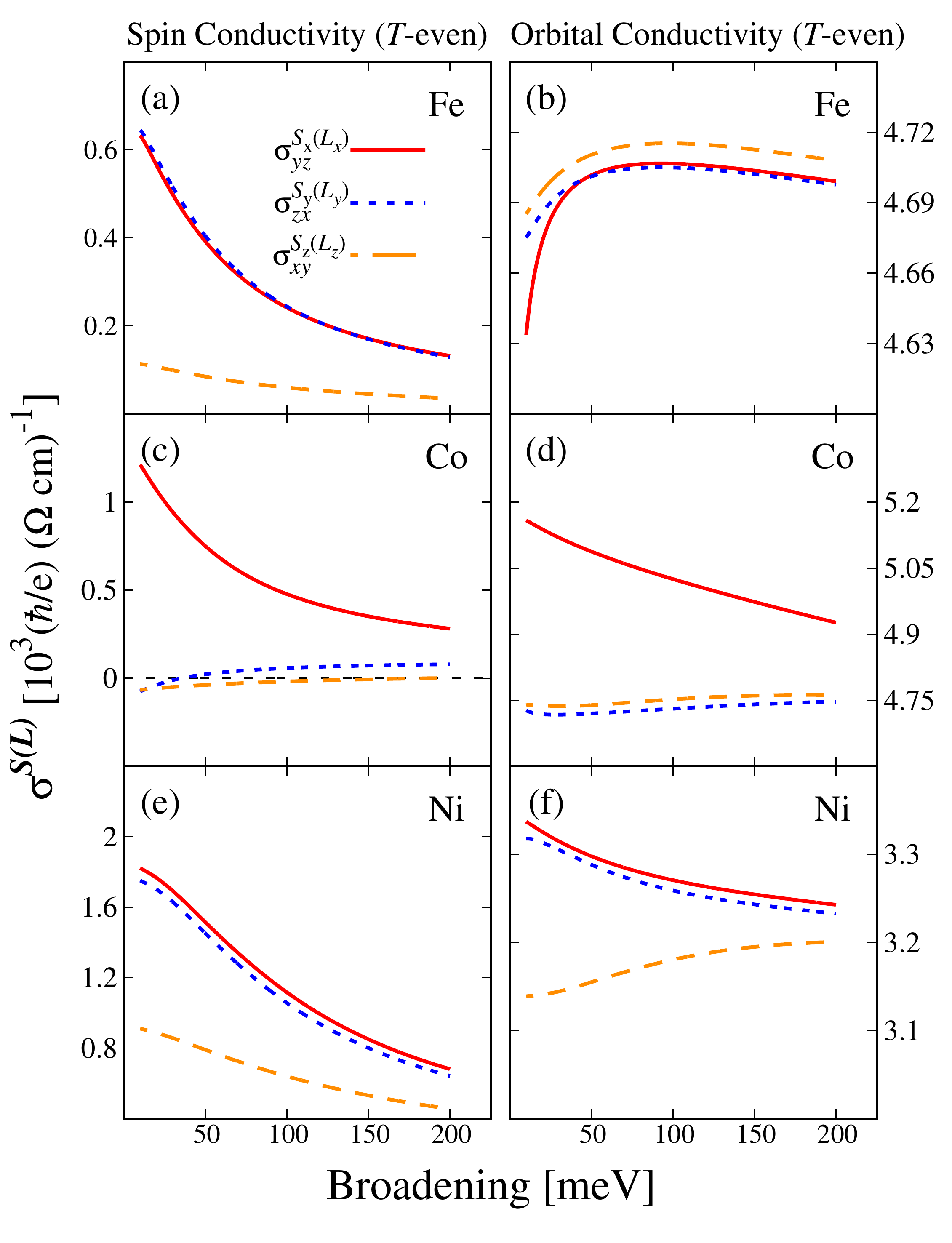}
  \caption{Dependence of the $\mathcal{T}$-even components on the lifetime parameter $\delta$ at $E=0$ for the nonzero elements of $\bm{\sigma}^{\bm{S}}$ for (a) Fe, (c) Co, and (e) Ni, and of $\bm{\sigma}^{\bm{L}}$ for (b) Fe, (d) Co, and (f) Ni. The $\bm{\sigma}^{\bm{L}}$ values vary at most a few percent with $\delta$, while the $\bm{\sigma}^{\bm{S}}$ components vary in much greater proportion and can even double when from $\hbar \delta=200$ meV to $10$ meV.}
  \label{fig:SHE-like-Conductivities-Broadening}
\end{figure}

Next, we investigate the lifetime dependency of both the SHE and OHE components. Calculated results for their dependence on the broadening $\hbar \delta$ is shown in Fig.~\ref{fig:SHE-like-Conductivities-Broadening}. It is important to note that both $\mathcal{T}$-even effects originate from the interband term, the intraband term vanishes. For the SHE, a significant dependence on the lifetime broadening is observed. Decreasing $\hbar \tau^{-1}$ from 200 meV to $10$ meV increases $\sigma^{S_x}_{yz}/\sigma^{S_y}_{zx}$ by $+380 \%$ and $\sigma^{S_z}_{xy}$ by $+220 \%$ for Fe, while those numbers are $+170 \%$ and $+100 \%$ for Ni. The case of Co is a bit different, yet $\sigma^{S_x}_{yz}$ shows an increase of $+330 \%$, similarly to what is observed for Fe and Ni. The two other components, $\sigma^{S_y}_{zx}$ and $\sigma^{S_z}_{xy}$, stay really close to $0$, with a sign inversion. At this point we can furthermore compare with the calculated intrinsic anomalous Hall conductivities of the ferromagnets, shown in detail in Appendix \ref{Electrical-conductivity}. This comparison exemplifies that the spin Hall and anomalous Hall conductivities can have opposite signs.
In contrast to the SHE,  for the OHE (right-hand column in Fig.~\ref{fig:SHE-like-Conductivities-Broadening}) the variation of the OHE with $\delta$ is practically negligible, typically within $1-2 \%$. This difference can already be understood from the sharply structured spectra of the SHE components, shown in Fig.\ \ref{fig:SHE-like-Conductivities}. These display moreover both positive and negative spectral peaks that will become reduced for a larger lifetime broadening, in contrast to the OHE spectra that are more smooth and always positive. 

\subsection{Magnetic spin and orbital Hall effect ($\mathcal{T}$-odd)}

We now focus on the MSHE and MOHE, which we respectively define as the electrical generation of a spin and an orbital current  arising from the $\mathcal{T}$-odd components of $\bm{\sigma}^{\bm{S}}$ and $\bm{\sigma}^{\bm{L}}$, with the exception of the diagonal elements of $\bm{\sigma}^{S_z}$ in Eq.\ (\ref{eq:sigma-Sz-cubic-explicit}).
These are the components  $\sigma^{S_x(L_x)}_{xz}$, $\sigma^{S_x(L_x)}_{zx}$, $\sigma^{S_y(L_y)}_{yz}$, $\sigma^{S_y(L_y)}_{zy}$. By symmetry, the $x$ and $y$ indices can be interchanged, that is $\sigma^{S_x}_{xz} = \sigma^{S_y}_{yz}$ and $\sigma^{S_x}_{zx} = \sigma^{S_y}_{zy}$, leaving us with two independent components.

\begin{figure}[tb!]
  \includegraphics[width=\linewidth]{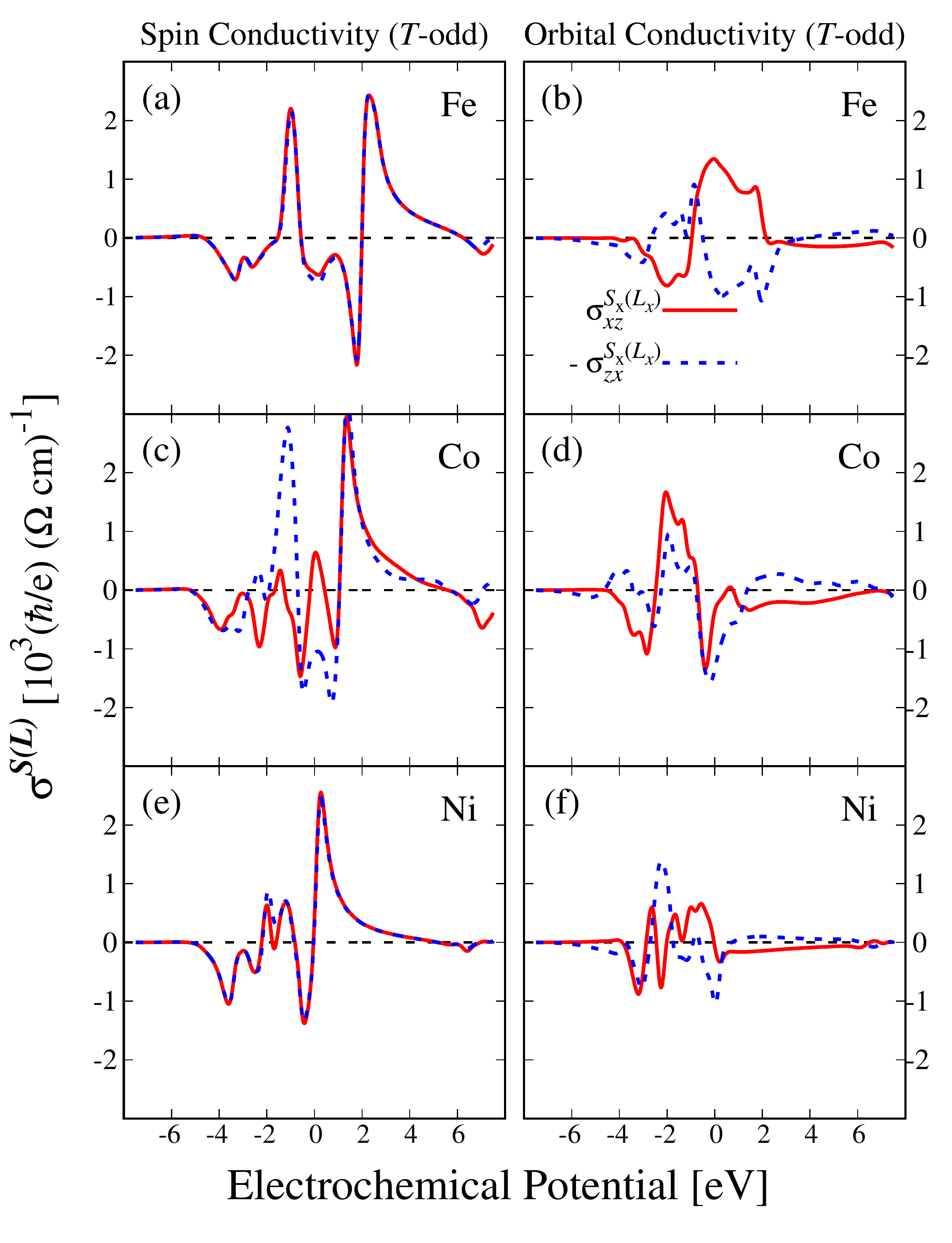}
  \caption{Calculated $\mathcal{T}$-odd components of $\bm{\sigma}^{\bm{S}}$ for (a) Fe, (c) Co, and (e) Ni, and of $\bm{\sigma}^{\bm{L}}$ for (b) Fe, (d) Co, and (f) Ni. The lifetime broadening used is $\hbar \tau^{-1} = 40$ meV. For sake of clarity $\sigma_{xz}^{S_x(L_x)}$ and $-\sigma_{zx}^{S_x(L_x)}$ are shown.}
  \label{fig:MSHE-like-Conductivities}
\end{figure}

Computed results for these components are shown in Fig.\ \ref{fig:MSHE-like-Conductivities}. For the MSHE (left-hand column of Fig.\ \ref{fig:MSHE-like-Conductivities}), we notice that the order of magnitude is similar to that of the $\mathcal{T}$-even SHE components. For Fe and Ni, we observe that $\sigma^{S_x}_{xz}$ and $-\sigma^{S_x}_{zx}$ are quite similar, while this doesn't hold for Co. Here again, the structural asymmetry due to the hcp lattice outweighs the magnetic asymmetry.
The MOHE components (right-hand column of Fig.\ \ref{fig:MSHE-like-Conductivities}) are of the same order of magnitude as the MSHE and SHE components. Also, in contrast to the MSHE,  $\sigma^{L_x}_{xz}$ is quite different from $-\sigma^{L_x}_{zx}$ for all three materials. The MSHE and MOHE conductivities furthermore display rather sharp spectral features, with both positive and negative peaks, in contrast to the larger OHE components shown in  Fig.\ \ref{fig:SHE-like-Conductivities}. The origin of this difference stems from the fact that nonzero OHE components are present even without SOC, but the SHE, MSHE, and MOHE components are induced by SOC.

The computed values for the MSHE and MOHE components at the Fermi level are given in Table~\ref{tab:SHE-OHE-Conductivities}. Although the MSHE components are rarely considered, they are comparably large as the SHE components for Fe and Co. For Ni the MSHE components are about four times smaller than the SHE values. The position of the Fermi energy for Ni plays a role in this difference.  $\sigma^{S_x}_{xz}$ changes steeply, from $-1371$ $\frac{\hbar}{e} (\Omega \,\text{cm})^{-1}$ at $E=-0.4$ eV to $2382.1$ $\frac{\hbar}{e} (\Omega \, \text{cm})^{-1}$ at $E=0.2$ eV. Similar sharp variations in the spectrum happens over $\sim 500$ meV  can be observed for Fe and Co, too, but the steep change is not at the Fermi energy.

\begin{figure}[bt!]
  \includegraphics[width=\linewidth]{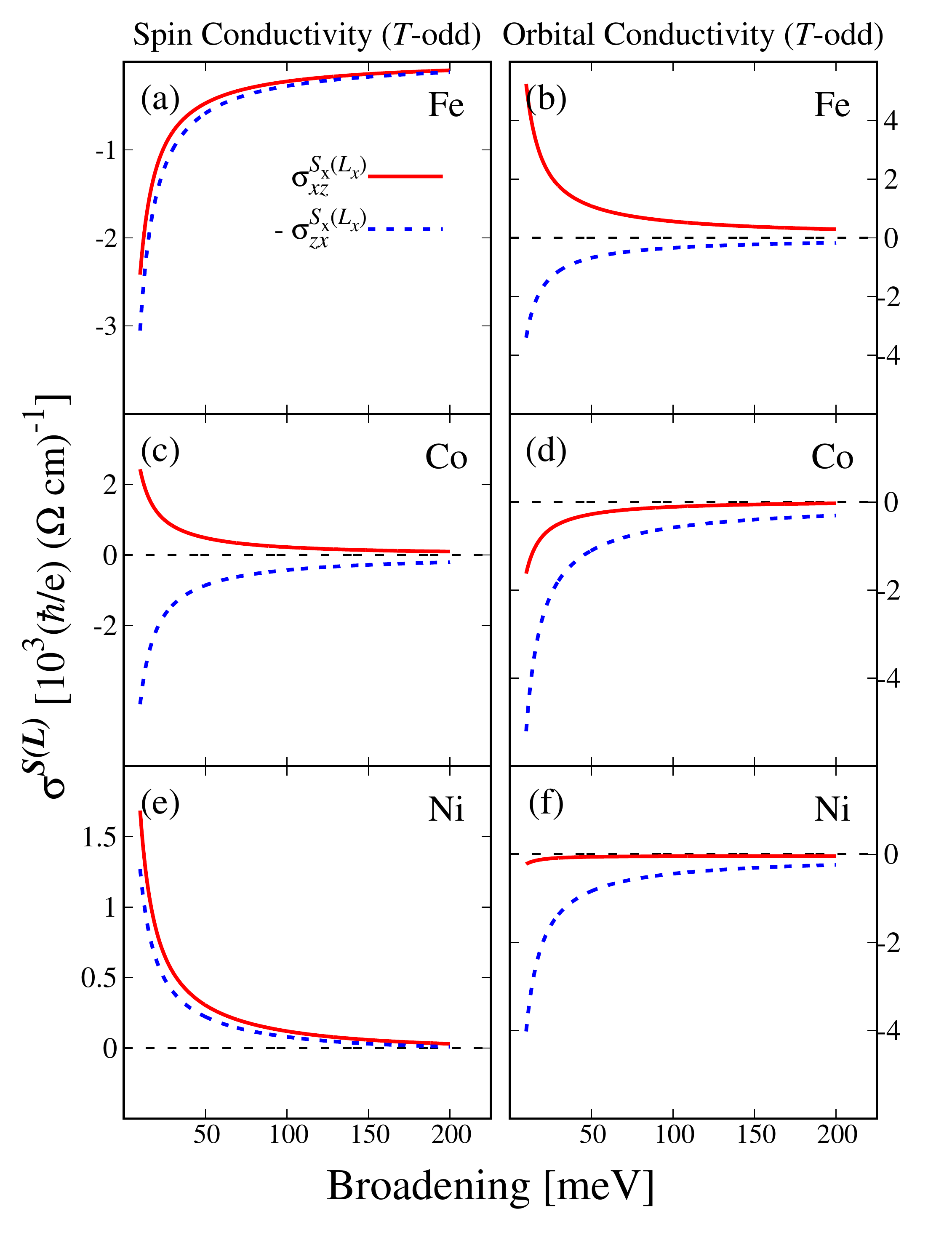}
  \caption{Lifetime-broadening dependence of the $\mathcal{T}$-odd  components of $\bm{\sigma}^{\bm{S}}$ for (a) Fe, (c) Co, and (e) Ni, and of $\bm{\sigma}^{\bm{L}}$ for (b) Fe, (d) Co, and (f) Ni, computed at $E=0$. The $\mathcal{T}$-odd components are mainly due to the intraband part of the response, therefore they all scale as $\propto \delta^{-1}$.}
  \label{fig:MSHE-like-Conductivities-Broadening}
\end{figure}

When it comes to the broadening dependence of the MSHE and MOHE, a completely different behavior than the SHE/OHE is observed, as both the MSHE and MOHE are intraband dominated effects. In Fig.\ \ref{fig:MSHE-like-Conductivities-Broadening} we show the lifetime dependence of the $\mathcal{T}$-odd components. As can be recognized, they do indeed scale as $\propto \tau^{-1}$. 
This has two fundamental implications for the MSHE and MOHE. First, contrarily to the SHE/OHE, the magnetic effects are theoretically unbounded. In ultra-clean samples, where the electron lifetime tends to increase, the MSHE and MOHE will become gigantic.
The MSHE components will then be larger than the SHE. 
Second, for dirty samples, or in the limit of large lifetime broadening, both the MSHE and MOHE become small. As there however remains an interband contribution to the MSHE and MOHE, the tensor components do not vanish and an explicit comparison with the values of the SHE and OHE is required.

\begin{figure}[ht!]
  \includegraphics[width=\linewidth]{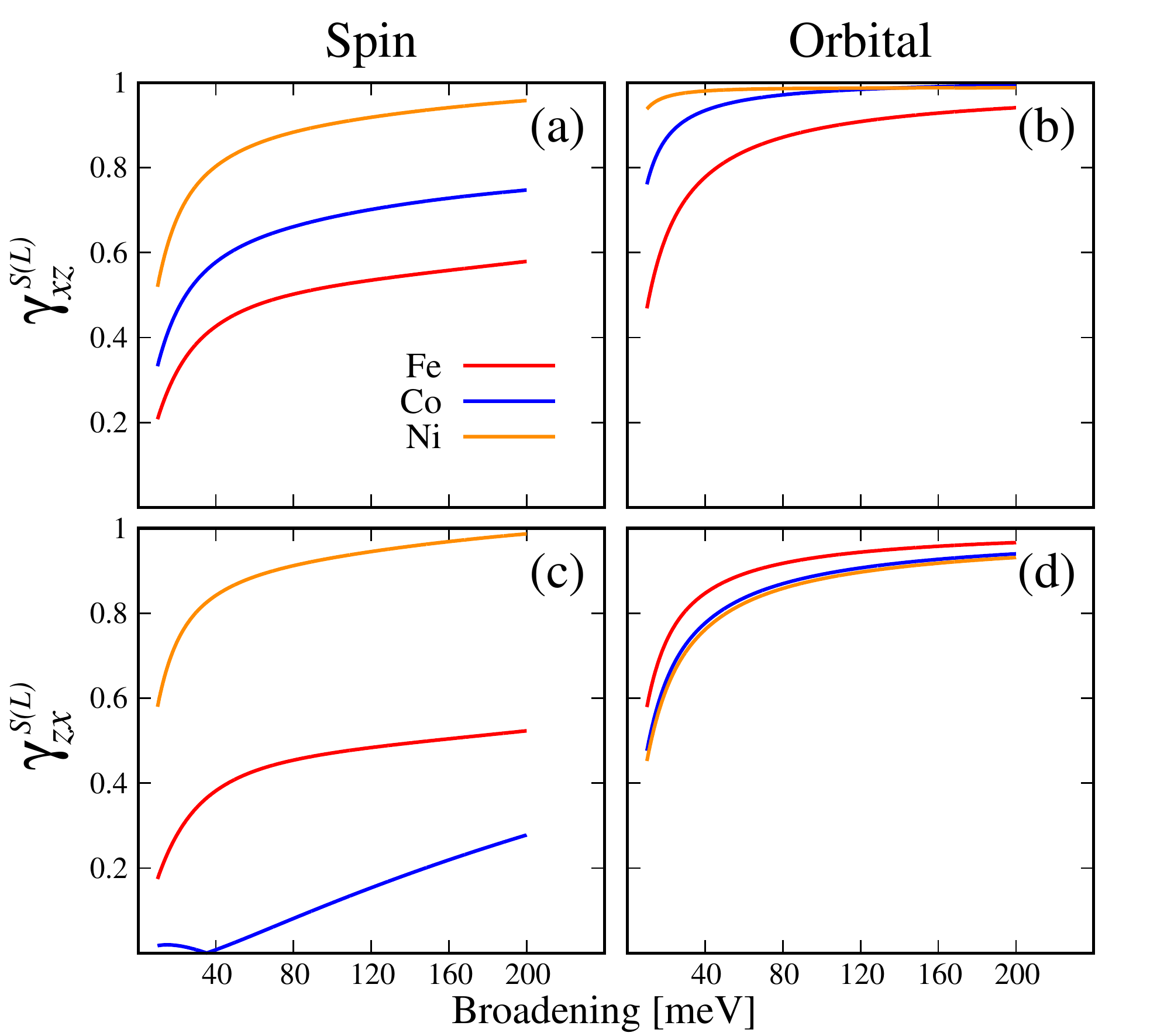}
  \caption{Ratios between the SHE and MSHE, and the OHE and MOHE, as defined in Eq.\ (\ref{eq:ratio}), as a function of the lifetime broadening. A value close to 1 (0) means that the spin/ orbital current originates mainly from the $\mathcal{T}$-even ($\mathcal{T}$-odd) SHE/OHE (MSHE/MOHE).}
  \label{fig:Ratios}
\end{figure}

To perform such comparison we define the ratio
\begin{equation}
\label{eq:ratio}
\gamma^{S(L)}_{ij} = \frac{|\sigma^{S_y(L_y)}_{ij}|}{|\sigma^{S_x(L_x)}_{ij}| + |\sigma^{S_y(L_y)}_{ij}|}  ,
\end{equation}
where $ij$ is either $xz$ or $zx$. When $\gamma^{S(L)}_{ij}$ approaches $1$, the SHE (OHE) is dominant over the MSHE (MOHE) while the opposite is true for $\gamma^{S(L)}_{ij} \rightarrow 0$. The calculated ratios $\gamma^{S(L)}_{xz}$ and $\gamma^{S(L)}_{zx}$ are displayed in Fig.\ \ref{fig:Ratios}. It is evident that all ratios increase with larger lifetime broadening $\delta$, implying that the SHE and OHE become dominant over the MSHE and MOHE, respectively. The ratio for the orbital effects is dominated by the OHE, which hardly changes with the lifetime broadening. However, for small broadenings the MSHE becomes larger than the SHE and will dominate over the SHE. This happens strongly for the $\gamma^{S}_{zx}$ ratio of Co. As a general trend, it can be seen that the larger the ferromagnetic moment is, the more dominant the MSHE is.

\subsection{Thermally-driven spin and orbital transport}

The generation of spin and orbital currents due to both an external electric field and a thermal gradient can be expressed as
\begin{align}
\label{eq:elec_thermal_spin_transport}
J^{S_k}_i &= \sigma^{S_k}_{ij} E_j - \Lambda^{S_k}_{ij}  \frac{dT}{dr_j},\\
\label{eq:elec_thermal_orbital_transport}
J^{L_k}_i &= \sigma^{L_k}_{ij} E_j - \Lambda^{L_k}_{ij}  \frac{dT}{dr_j},
\end{align}
where $\Lambda^{S_k (L_k)}_{ij}$ is the spin (orbital) magnetothermal conductivity tensor.  
These thermal transport tensors $\bm{\Lambda}$ can be extracted from $\bm{\sigma}^{\bm{S}}(E)$ and $\bm{\sigma}^{\bm{L}}(E)$ using the Mott equation (\ref{eq:Mott_Formula}). Again, our focus is here on the transverse coefficients, $\bm{\Lambda}^{S(L)}$ specifically, the spin Nernst effect (SNE) and magnetic spin Nernst effect (MSNE) and the orbital Nernst effect and magnetic orbital Nernst effect (ONE and MONE).

Results for the calculated SNE, MSNE and their orbital  counterparts as function of the electrochemical potential $E$ are given in Appendix \ref{thermal-conductivities} (Fig.\ \ref{fig:Thermal-Conductivities}). It can be noted that these spin and orbital thermal conductivities depend significantly on the chemical potential. Features comparable to those of the SHE and MSHE and their orbital counterparts can be observed: The magnetic spin and orbital thermal conductivities ($\mathcal{T}$-odd) are similarly large as the nonmagnetic ($\mathcal{T}$-even) conductivities and the orbital thermal conductivity is very isotropic. 

\begin{figure*}[ht!]
  \includegraphics[width=0.90\linewidth]{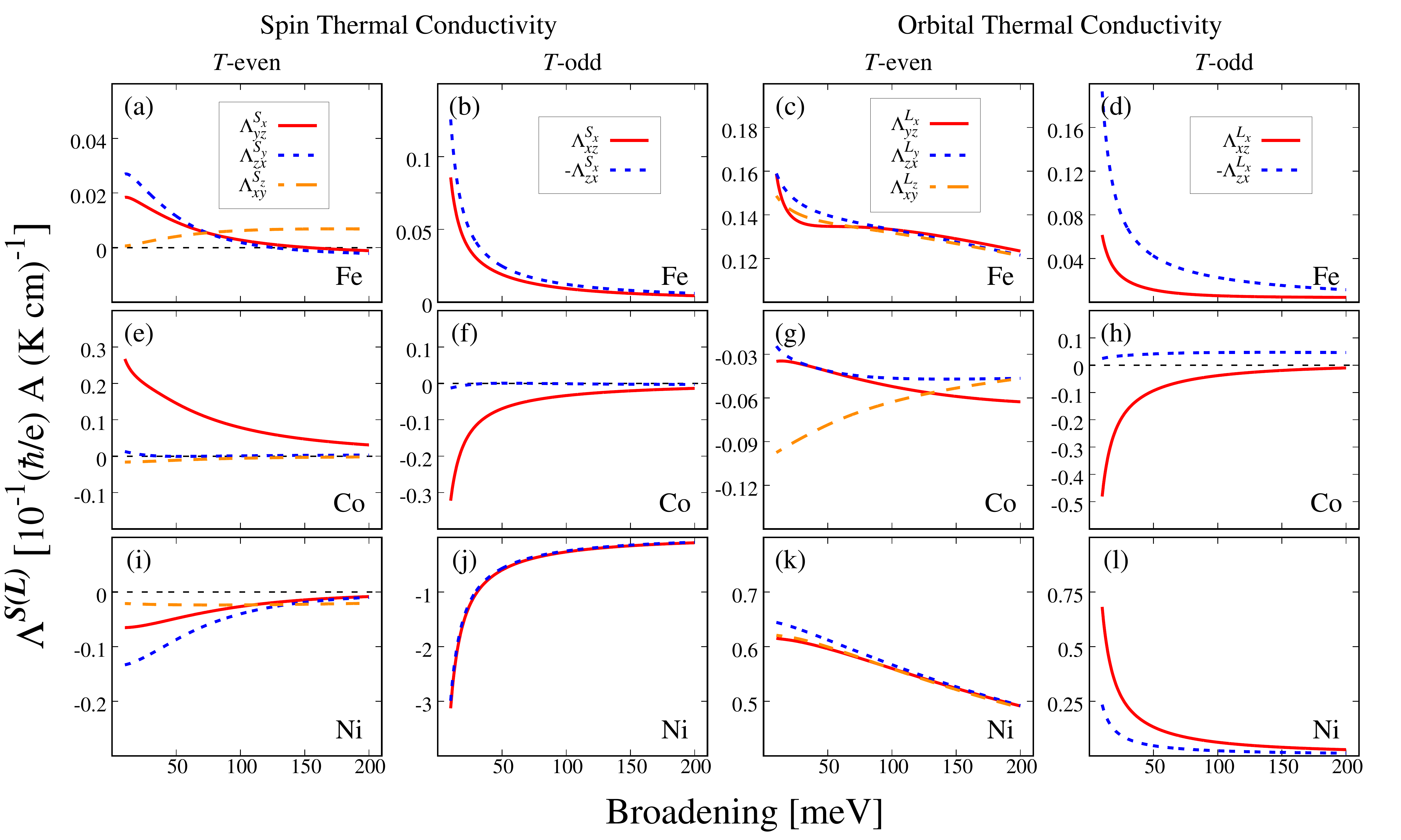}
  \caption{Calculated spin and orbital transverse thermal conductivities in Fe, Co, and Ni as a function of the lifetime broadening $\hbar \delta$ (at $E = 0 $ eV). Shown from left to right is the SNE, MSNE, ONE, and MONE.}
  \label{fig:Thermal-Conductivities-Broadening}
\end{figure*}

It is instructive to consider the dependence of the transverse thermal conductivities at the Fermi energy on the lifetime broadening $\hbar\delta$, shown in Fig.\ \ref{fig:Thermal-Conductivities-Broadening}. The magnetic spin and orbital thermal conductivities MSNE and MONE increase steeply as $\sim \delta^{-1}$ for small lifetime broadenings whereas the SNE and ONE approach stable values for small $\delta$.  The MSNE and MONE are clearly not negligible, they can be equally large or larger than the SNE and ONE in ultra-clean samples.

The Seebeck coefficient $S$ is commonly defined as the longitudinal thermal coefficient divided by the longitudinal charge conductance, $S= \Lambda_{ii}/ \sigma_{ii}$.  Similarly to the definition of the Seebeck coefficient, we can define transverse spin and orbital transport  coefficients $\alpha$  as
\begin{equation}
\alpha^{S_{k}(L_{k})}_{ij} = \frac{\Lambda^{S_{k}(L_{k})}_{ij}}{\sigma^{S_{k}(L_{k})}_{ij}},
\end{equation}
where $\alpha^{S_{k}(L_{k})}_{ij}$ is given in units of ${\text{V}\,}{\text{K}}^{-1}$. {For the materials considered in this paper, the coefficients $\alpha^{S_x}_{yz}$ ($\alpha^{L_x}_{yz}$), $\alpha^{S_y}_{zx}$ ($\alpha^{L_y}_{zx}$) and $\alpha^{S_z}_{xy}$ ($\alpha^{L_z}_{xy}$) quantify the SNE (ONE) and $\alpha^{S_x}_{xz}$ ($\alpha^{L_x}_{xz}$) and $\alpha^{S_x}_{zx}$ ($\alpha^{L_x}_{zx}$) the MSNE (MONE) with respect to the  corresponding electrical spin and orbital conductivities. We shall refer to those coefficients as spin Nernst coefficient (SNC),  magnetic spin Nernst coefficient (MSNC), orbital Nernst coefficient (ONC), and magnetic orbital Nernst coefficient (MONC).}

The results of our calculations are summarized in Table~\ref{tab:spin-thermal-transport} for the SNC and MSNC and Table~\ref{tab:orbital-thermal-transport} for the ONC and MONC. These transport coefficients are computed at the Fermi energy ($E=0$ eV), for $T=300$ K and $\tau^{-1} = 40$ meV.

\begin{table}[h!]
\begin{ruledtabular}
\caption{Calculated transverse spin thermal transport coefficients  $\bm{\alpha}^{\bm{S}}$, for ferromagnetic bcc Fe, hcp Co, and fcc Ni. The SNE is quantified by $\alpha^{S_x}_{yz}$, $\alpha^{S_y}_{zx}$, and $\alpha^{S_z}_{xy}$ while the MSNE by $\alpha^{S_x}_{xz}$ and $\alpha^{S_x}_{zx}$. The thermal transport coefficients are given in ${\mu \text{V}}{\text{K}^{-1}}$  and for $T = 300$\,K.}
\label{tab:spin-thermal-transport}
\centering
\begin{tabular}{l ccc c cc}
 & \multicolumn{3}{c}{SNC} & & \multicolumn{2}{c}{MSNC} \\
 \cline{2-4} \cline{6-7} \tabularnewline[-3\doublerulesep] 
  & $\alpha^{S_x}_{yz}$ & $\alpha^{S_y}_{zx}$ & $\alpha^{S_z}_{xy}$ & & $\alpha^{S_x}_{xz}$ & $\alpha^{S_x}_{zx}$ 
       \tabularnewline[\doublerulesep]
    \hline     
         \tabularnewline[-3\doublerulesep] 
Fe & 2.63 & 3.32 & 3.57 & & -3.94 & -4.18 \\[0.35em]
Co & 19.72 & 1.10 & 28.74 & & -14.12 & ~2.91 \\[0.35em]
Ni & -3.35 & -6.50 & -2.80 & & -192.55 & -248.60 \\
\end{tabular}
\end{ruledtabular}
\end{table}

\begin{table}[h!]
\begin{ruledtabular}
\caption{As Table \ref{tab:spin-thermal-transport}, but for the transverse orbital thermal transport coefficients $\bm{\alpha}^{\bm{L}}$, specifically, the ONC 
($\alpha^{L_x}_{yz}$, $\alpha^{L_y}_{zx}$, and $\alpha^{L_z}_{xy}$) and the MONC ($\alpha^{L_x}_{xz}$ and $\alpha^{L_x}_{zx}$).}
\label{tab:orbital-thermal-transport}
\centering
\begin{tabular}{lccc c cc}
& \multicolumn{3}{c}{ONC} & & \multicolumn{2}{c}{MONC} \\
 \cline{2-4} \cline{6-7} \tabularnewline[-3\doublerulesep] 
  & $\alpha^{L_x}_{yz}$ & $\alpha^{L_y}_{zx}$ & $\alpha^{L_z}_{xy}$ & & $\alpha^{L_x}_{xz}$ & $\alpha^{L_x}_{zx}$ 
    \tabularnewline[\doublerulesep]
    \hline     
         \tabularnewline[-3\doublerulesep] 
Fe & 2.87 & 3.02 & 2.93 & & 1.07 & -6.10 \\[0.35em]
Co & -0.77 & -0.83 & -1.75 & & 33.48 & ~9.67 \\[0.35em]
Ni & 18.22 & 18.85 & 19.26 & & -253.57 & -5.61 \\
\end{tabular}
\end{ruledtabular}
\end{table}

Looking at the spin $\alpha$'s, we see that in the case of Fe they are of similar magnitude, that is, $\alpha \sim 3 \, {\mu \text{V}}{\text{K}}^{-1}$ for the SNC, and $\alpha \sim -4 \,{\mu \text{V}}{\text{K}}^{-1}$ for the MSNC. For Co, a strong anisotropy is observed, for the SNC with $\alpha^{S_x}_{yz}, ~ \alpha^{S_z}_{xy}  > \alpha^{S_y}_{zx}$, and for the MSNC with $|\alpha^{S_x}_{zx}|  > |\alpha^{S_x}_{xz}|$ where the absolute value is taken because the signs are opposite. While both Fe and Co have SNCs of the same magnitude as the MSNCs, we observe that this is not the case for Ni. Remarkably, the MSNCs in Ni are 2 orders of magnitude higher than the SNCs, with all coefficients being of the same sign.

For the ONC and MONC, the comparison to their spin counterpart depends strongly on the material considered. For Fe, the SNC and ONC are remarkably close. For Co, the ONCs and MONCs are respectively smaller and bigger than their spin counter part. For Ni, the ONCs are $3$ to $6$ times larger than their spin counter part. The anisotropy for the ONC is virtually non-existant, even in the case of Co which had a strong structure-induced anisotropy in its OHE components. While so far the spin Nernst effect has been observed only in Pt and W \cite{Meyer2017,Sheng2017,Bose2018}
our calculations suggest that in particular it should be possible to measure a large unusual MSNE in Ni, being much larger than the same of Fe and Co.

\section{Discussion}
\label{Discussion}

\subsection{Charge-to-spin Conversion}
Theoretical investigations usually discuss spin transport on the basis of the spin conductivity tensor $\bm{\sigma}^{\bm{S}}$, because the influence of the external perturbation is described directly in terms of the electric field $\bm{E}$. On the other hand, experimental works focus on the conversion of a charge-current density $\bm{J}$ to an output spin-current density $\bm{J}^{\bm{S}}$. Here, we will discuss these two pictures, and relate them.

We start by considering a charge current density $\bm{J}$. In the linear regime
\begin{equation}
\label{eq:chargecurrent-electricfield-relation}
\bm{J} = \bm{\sigma} \bm{E},
\end{equation}
where $\bm{\sigma}$ is the electrical conductivity tensor. Combining Eqs.\ (\ref{eq:spincurrent-electricfield-relation}) and  (\ref{eq:chargecurrent-electricfield-relation}), we can write
\begin{equation}
\label{current-spincurrent-relation}
\bm{J}^{S_k} = \bm{\sigma}^{S_k} \bm{\sigma}^{-1}  \, \bm{J} 
= \bm{\sigma}^{S_k} \bm{\rho}  \, \bm{J} 
= \frac{\hbar}{2e} \bm{\theta}^{S_k}   \, \bm{J} ,
\end{equation}
where $\bm{\rho} = \bm{\sigma}^{-1}$ is the resistivity tensor and $\bm{\theta}^{\bm{S}}$ is a 3$^{\text{rd}}$ rank tensor which is the generalization of the concept of the spin Hall angle ({SHA}). Note that the element $\theta^{S_z}_{yx}$ would be the commonly defined SHA for nonmagnetic metals. We will refer to $\bm{\theta}^{\bm{S}}$ as the \emph{spin-charge angle} (SCA) tensor.

The resistivity tensor $\bm{\rho}$ can explicitly be written as
\begin{equation}
\label{eq:rho-cubic-explicit}
\bm{\rho} =
\begin{pmatrix}
\rho_1 & \rho_A & 0 \\
-\rho_A & \rho_1 & 0 \\
0 & 0 & \rho_2 \\
\end{pmatrix}
,
\end{equation}
where $\rho_1$ and $\rho_2$ are the $\mathcal{T}$-even diagonal part of $\bm{\rho}$ and $\rho_A$ is the anomalous Hall resistivity, which is $\mathcal{T}$-odd. The elements $\sigma_{ij}$ of the conductivity tensor can be computed with the same linear-response formulation. Results for the ferromagnetic elements are given in Appendix \ref{Electrical-conductivity}. Inserting Eqs.\ (\ref{eq:sigma-Sx-cubic-explicit}), (\ref{eq:sigma-Sy-cubic-explicit}), (\ref{eq:sigma-Sz-cubic-explicit}), and (\ref{eq:rho-cubic-explicit}) in Eq.\ (\ref{current-spincurrent-relation}) we can find an explicit expression of $\bm{\theta}^{\bm{S}}$
\begin{subequations}
\scriptsize
\begin{align}
\label{eq:theta-Sx-cubic-explicit}
\bm{\theta}^{S_x} &= \frac{2e}{\hbar}
\begin{pmatrix}
0 & 0 & { \sigma^{S_x}_{xz}} \rho_2 \\
0 & 0 & {\blue \sigma^{S_x}_{yz} \rho_2 } \\
{ \sigma^{S_x}_{zx}} \rho_1 - { \sigma^{S_x}_{zy}} \rho_A  & { \sigma^{S_x}_{zx}} \rho_A + {\blue \sigma^{S_x}_{zy} \rho_1 } & 0 \\
\end{pmatrix}
,\\
\label{eq:theta-Sy-cubic-explicit}
\bm{\theta}^{S_y} &= \frac{2e}{\hbar}
\begin{pmatrix}
0 & 0 & {\blue \sigma^{S_y}_{xz} \rho_2} \\
0 & 0 & { \sigma^{S_y}_{yz}} \rho_2 \\
{\blue \sigma^{S_y}_{zx} \rho_1} - { \sigma^{S_y}_{zy}} \rho_A  & { \sigma^{S_y}_{zx}} \rho_A + { \sigma^{S_y}_{zy}} \rho_1 & 0 \\
\end{pmatrix}
,\\
\label{eq:theta-Sz-cubic-explicit}
\bm{\theta}^{S_z} &= \frac{2e}{\hbar}
\begin{pmatrix}
\sigma^{S_z}_{xx} \rho_1 - { \sigma^{S_z}_{xy}} \rho_A & \sigma^{S_z}_{xx} \rho_A + {\blue \sigma^{S_z}_{xy} \rho_1} & 0\\
{\blue \sigma^{S_z}_{yx} \rho_1} - \sigma^{S_z}_{yy} \rho_A & { \sigma^{S_z}_{yx}} \rho_A + \sigma^{S_z}_{yy} \rho_1 & 0 \\
0 & 0 & \sigma^{S_z}_{zz} \rho_2 \\
\end{pmatrix}.
\end{align}
\normalsize
\end{subequations}
These expressions form a bridge between the theoretical ``$\bm{E}$-in $\bm{J}^{\bm{S}}$-out" and experimental ``$\bm{J}$-in $\bm{J}^{\bm{S}}$-out" picture. The conventional, nonmagnetic SHE elements are indicated with blue color. Compared to $\bm{\sigma}^{\bm{S}}$, $\bm{\theta}^{\bm{S}}$ shows additionally a more complex structure because of the mixing of tensor components.

First, let us look at the SHE-like components, that is, $\theta^{S_k}_{ij}$ where $\epsilon_{ijk} \neq 0$. Depending on the orientation of the spin polarization $\bm{S}$, of the spin current $\bm{J}^{\bm{S}}$, and of the charge current $\bm{J}$ relative to $\bm{M}$, we can classify those components as
\begin{itemize}
\item[-] $\theta^{S_x}_{yz}$ and $\theta^{S_y}_{xz}$: $\bm{S} \bot \bm{M}$ and $\bm{M} \parallel \bm{J}$,
\item[-] $\theta^{S_x}_{zy}$ and $\theta^{S_y}_{zx}$: $\bm{S} \bot \bm{M}$ and $\bm{M} \bot \bm{J}$,
\item[-] $\theta^{S_z}_{xy}$ and $\theta^{S_z}_{yx}$: $\bm{S} \parallel \bm{M}$ and $\bm{M} \bot \bm{J}$.
\end{itemize}
Although those components are all SHE-like, their physical interpretation differs greatly. The component $\theta^{S_x}_{yz} =
\sigma_{yz}^{S_x} \rho_2$ ($\theta^{S_y}_{xz}= \sigma_{xz}^{S_y} \rho_2$) emerges from the interplay of the spin conductivity tensor element $\sigma^{S_x}_{yz}$ ($\sigma^{S_y}_{xz}$) and the longitudinal-to-$\bm{J}$ potential gradient $\partial_z V \sim \rho_2 J_z$. This component can be understood as the simple extension of $\sigma^{S_x}_{yz}$ ($\sigma^{S_y}_{xz}$) in a SCA perspective.

The component $\theta^{S_x}_{zy} = \sigma_{zx}^{S_x} \rho_A+ \sigma_{zy}^{S_x} \rho_1$, {which is not} symmetrical to $\theta^{S_x}_{yz}$, shows a more complex structure. It is expressed as the sum of two terms: (1) $\sigma^{S_x}_{zx} \rho_A$ and (2) $\sigma^{S_x}_{zy} \rho_1$. While the physical picture of the second term is analogous to what has been discussed in the previous paragraph, the first term is different. It can be interpreted as the following. Because of the AHE, an external charge current $J_y$ produces a transverse potential gradient $\partial_x V \sim \rho_A J_y$, which gives rise to a spin current due to the MSHE-like spin conductivity $\sigma^{S_x}_{zx}$. Interestingly enough, both $\sigma^{S_x}_{zx}$ and $\rho_A$ are $\mathcal{T}$-odd, but because it is their product that comes into play, this term is experimentally indistinguishable from a conventional SHE-generated spin-current, i.e., it possesses the same spacial and time-reversal symmetries. A similar discussion can be held for $\theta^{S_y}_{zx}$.

The remaining two SHE-like components are $\theta^{S_z}_{xy}$ and $\theta^{S_z}_{yz}$. Those components tend to be referred to as SHA, although it should be clear by now that defining a unique value for the SHA can be misleading in lower symmetry systems. We focus on $\theta^{S_z}_{xy}$, since the discussion of $\theta^{S_z}_{yx}$ is similar. The component $\theta^{S_z}_{xy}$ is written as $\theta^{S_z}_{xy} = \sigma^{S_z}_{xx} \rho_A + \sigma^{S_z}_{xy} \rho_1$. Similarly to the other SHE-like components, the second term can be understood as the extension of the SHE-like spin conductivity $\sigma^{S_z}_{xy}$ in a SCA perspective. The first term is however peculiar, and can be interpreted as the following. An external charge current $J_y$ produces a AHE-induced transverse potential gradient $\partial_x V \sim \rho_A J_y$. Because the material is ferromagnetic, the longitudinal conductivity of spin up and spin down electron is different (spin filtering), i.e., $\sigma^{S_z}_{xx} \neq 0$, and therefore the current induced by $\partial_x V$ is inherently spin polarized. This contribution has drawn attention recently and has been discussed in terms of the spin anomalous Hall effect \cite{Das2017,Bose2018b,Gibbons2018,Iihama2018,Seki2019}. It is an $\mathcal{T}$-even effect and  experimentally indistinguishable from a conventional SHE-generated spin current.

Next, we discuss the MSHE-like components, that are the remaining components from $\bm{\theta}^{S_x}$ and $\bm{\theta}^{S_y}$, i.e., $\theta^{S_x}_{xz}$, $\theta^{S_x}_{zx}$, $\theta^{S_y}_{yz}$, and $\theta^{S_y}_{zy}$. All those components are $\mathcal{T}$-odd, with their spin-polarization direction orthogonal to $\bm{M}$. We can distinguish two cases. First, if $\bm{J}$ is parallel to $\bm{M}$, we have $\theta^{S_x}_{xz}$ and $\theta^{S_y}_{yz}$. In this case, the SCA components can be understood as the generation of spin current due to the MSHC, expressed in a ``$\bm{J}$-in $\bm{J}^{\bm{S}}$-out" picture. Remarkably, the spin-polarization direction is parallel to the direction of the flow of the spin current, which cannot be obtained with the SHE.

The second case is when $\bm{J}$ is orthogonal to $\bm{M}$ and $\bm{J}^{\bm{S}}$ parallel to $\bm{M}$, that is, $\theta^{S_x}_{zx}$ and $\theta^{S_y}_{zy}$. If we look at $\theta^{S_x}_{zx}$ ($\theta^{S_y}_{zy}$ is analogous), we see that it is expressed as the sum of two components: (1) $\sigma^{S_x}_{zx} \rho_1$ and (2) $- \sigma^{S_x}_{zy} \rho_A$. The first term is, as for $\theta^{S_x}_{xz}$, the generation of spin current due to the MSHC, expressed in a ``$\bm{J}$-in $\bm{J}^{\bm{S}}$-out" picture. For the second term, the physical picture is the following. A current $J_x$ produced a transverse potential gradient $\partial_y V \sim J_x \rho_A$ because of the AHE, which creates a spin current due to the SHE-like spin conductivity $\sigma^{S_x}_{zy}$. Although $\sigma^{S_x}_{zy}$ is $\mathcal{T}$-even, because it is driven by the AHE which is $\mathcal{T}$-odd, the effect is $\mathcal{T}$-odd. Here, the spin polarization direction is parallel to the direction of the input charge current, which cannot be obtained with the SHE.

The last group of components we discuss are the diagonal components of $\bm{\theta}^{S_z}$. For $\theta^{S_z}_{zz}$, the picture is simple, with $\theta^{S_z}_zz = \sigma^{S_z}_{zz} \rho_2$ showing the extension of conductivity in the SCA picture. For $\theta^{S_z}_{xx}$ and $\theta^{S_z}_{yy}$, one of the terms that defines them is similar to $\theta^{S_z}_{zz}$. The other one comes from the interplay of the AHE and SHE. This is quite interesting as it shows that a longitudinal charge current creates a longitudinal spin current, not only because of the difference of spin up and spin down conductivity, but also due to transverse SHE.

\subsection{Relation to other work}

The existence of anomalous SHE terms has in principle been known since the group-theoretical symmetry analysis of Seemann \textit{et al.}\ \cite{seemannSymmetryimposedShapeLinear2015}. Also the existence of the orbital Hall effect has been predicted years ago \cite{kontaniStudyIntrinsicSpin2007,Tanaka2008}. Still, not much is known about the actual sizes of the unconventional ($\mathcal{T}$-odd) spin and orbital effects. Several recent works initiated recently a discussion of these unusual effects. Humphries \textit{et al.}\ \cite{humphriesObservationSpinorbitEffects2017} observed an unusual magnetization-direction dependent spin torque for ferromagnetic/nonmagnetic metal stack, which they explained with a magnetization-linear spin current.   Kimata \textit{et al.}\ \cite{kimataMagneticMagneticInverse2019} reported the observation of the MSHE for a noncollinear antiferromagnet, Mn$_3$Sn. Mook \textit{et al.}\ \cite{mookOriginMagneticSpin2020} analyzed the origin of the MSHE and attributed it to spin-current vorticity in the Fermi sea for the noncollinear antiferromagnet. For the ferromagnetic 3$d$ elements we find that the $\mathcal{T}$-odd MSHE components mainly originates from the intraband response contribution, i.e., from the Fermi surface. Salemi \textit{et al.}\ \cite{Salemi2021} investigated the MSHE and MOHE for ferromagnetic metal/Pt bilayer films and computed non-negligible MSHE conductivities in the ferromagnetic layer. 

In a recent paper, Qu \textit{et al.}\ \cite{quMagnetizationDirectionDependent2020} reported that they calculated a magnetization dependent SHC. However, as compared to our study, what they really computed was different $\mathcal{T}$-even tensor elements of $\bm{\sigma}^{\bm{S}}$ (specifically, its magnetocrystalline anisotropy). Those tensor elements should be equivalent in cubic system, but only when the magnetism is turned off. Miura and Masuda \cite{Miura2021} investigated the spin anomalous Hall effect for $X$Pt ($X$= Fe, Co, Ni) which is defined as the anisotropy of the $\mathcal{T}$-even elements when the magnetization is along the tetragonal $c$ axis or in the basal plane.

A thorough analysis of the spin currents that could appear in a ferromagnetic material was recently provided by Wang \cite{Wang2021}. 
The symmetry-allowed anomalous SHE tensor elements predicted by Wang are indeed fully confirmed by our calculations. A distinction is that in our formulation one can recognize the origin of an anomalous SCA tensor element, e.g., $\theta_{zx}^{S_x} = (2 e / \hbar )(\sigma_{zx}^{S_x} \rho_1 -\sigma_{zy}^{S_x} \rho_A)$, whereas in Wang's analysis it is an allowed nonzero element and because of the Onsager reciprocity, there will be a related inverse effect \cite{Wang2021}.

\section{Conclusions}
We have used first-principles calculations to investigate the electric and thermal generation of spin and orbital currents in the bulk ferromagnets Fe, Co, and Ni. For each material, we have computed all the nonzero components of the relevant tensors, that is, $\bm{\sigma}$, $\bm{\sigma}^{\bm{S}}$, $\bm{\sigma}^{\bm{L}}$, $\bm{\Lambda}^{\bm{S}}$, and $\bm{\Lambda}^{\bm{L}}$.

Our extensive study has shown that defining the SHC in lower symmetry systems is more involved than for nonmagnetic cubic materials like Pt, as the relative orientation of the $\bm{M}$ with respect to the electric field, the spin current, and spin polarization of the spin current plays a crucial role. This non-uniqueness in SHC can have lead to some confusion. 

We have shown that for the SHE, the spin conductivity from the tensor elements of $\bm{\sigma}^{\bm{S}}$ whose spin-polarization is perpendicular to the $\bm{M}$ direction tend to be several times larger than the ones where $\bm{S}$ is parallel to $\bm{M}$. This has quite important implications, since it is common in the field to focus on $\bm{S}$ and $\bm{M}$ along $\bm{u}_z$. Thorough investigation of different configurations for complex systems could lead to increased efficiency in charge-to-spin conversion.

We have also investigated the recently proposed MSHE, that is odd under time-reversal symmetry. We have computed the  \emph{ab initio} material dependent MSHE-like conductivities for the simple ferromagnets Fe, Co, and Ni. It turns out that those components are not only far from negligible, but actually on par with the SHE related component. Also, contrarily to the SHE, the MSHE has an intraband component, meaning that ultra-clean system should see a gigantic effect. We have computed a similar effect for the orbital part, that is the MOHE, that has not been proposed in the literature before.

As suggested by Mook \textit{et al.}\ \cite{mookOriginMagneticSpin2020}, because the MSHE exists, a thermal counter part must exist, too. This is the MSNE, which we have thoroughly investigated in this work. We have extended the concept of the magnetic spin Nernst effect to the orbital angular momentum, the MONE. We have evaluated those two effects, and investigated their dependency with respect to the electrochemical potential as well as the lifetime broadening. Observations of the spin Nernst effect are still scarce, but we hope that our first-principles calculations will stimulate investigations of the ONE and the MSNE and MONE.

Lastly, the prediction of sizeable MSHE and MOHE in magnetic materials could have some deep implications for device design. While the conventional SHE allows for an input charge current, output spin current and spin polarization that are all mutually orthogonal, the MSHE enables a more complex generation of a spin current, where two of those components can become parallel. This could be utilized in the design of special switching geometries for spintronics devices.

\section*{Acknowledgments}
This work has been funded by the European Union's Horizon2020 Research and Innovation Programme under FET-OPEN Grant agreement No.\ 863155 (s-Nebula). This work has furthermore been supported by the Swedish Research Council (VR).  The calculations were performed at the PDC Center for High Performance Computing and the National Supercomputer Center (NSC) enabled by resources provided by the Swedish National Infrastructure for Computing (SNIC), partially funded by the Swedish Research Council through grant agreement No.\ 2018-05973.

\appendix
\section{Electrical conductivity}
\label{Electrical-conductivity}

The relevant components of the electrical conductivity tensor $\bm{\sigma}$ can be computed as well using the linear-response formalism, Eq.\ (\ref{eq:LinearResponse}), but using the momentum $\hat{\bm{p}}$ instead of the spin current operator.

\begin{figure}[t!]
  \includegraphics[width=1.0\linewidth]{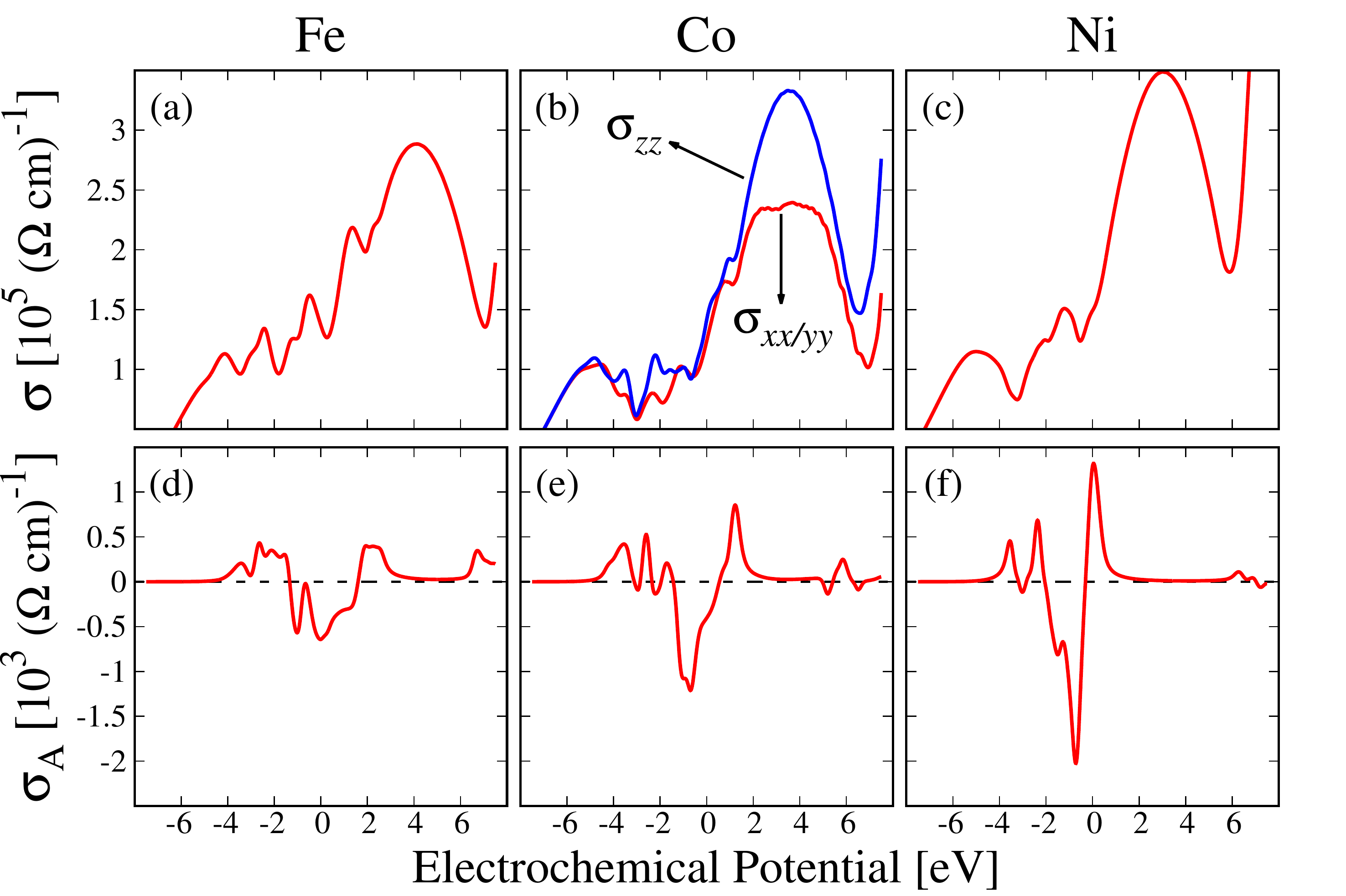}
  \caption{Calculated longitudinal conductivity $\sigma$ for (a) Fe, (b) Co, and (c) Ni as well as the anomalous Hall conductivity $\sigma_A$  for (d) Fe, (e) Co, and (f) Ni. For Co, because of the hexagonal structure, the in-plane $\ \sigma_{xx} = \sigma_{yy}$ conductivity and the out-of-plane $ \sigma_{zz}$ are distinct. The anomalous conductivity is given as $\sigma_A = \sigma_{xy} = -\sigma_{yx}$. The lifetime broadening used is $\hbar\delta = 40$ meV.}
  \label{fig:Conductivities}
\end{figure}

In Fig.\ \ref{fig:Conductivities}(a), (b), and (c), we give for completeness' sake the computed longitudinal conductivities $\sigma_1$ and $\sigma_2$, respectively, for Fe, Co, and Ni as a function of the electrochemical potential $E$. As discussed earlier, those are given by the diagonal elements of $\bm{\sigma}$. For the cubic materials Fe and Ni, the lowering of symmetry caused by $\bm{M}$ has negligible impact on the asymmetry between $\sigma_1$ and $\sigma_2$ (less than $1\%$ difference). For Co, the distinction between $\sigma_1$ and $\sigma_2$ must be taken into account because of the structural asymmetry of the hcp structure [see Fig.\ \ref{fig:Conductivities}(b)].

\begin{figure}[tb!]
  \includegraphics[width=0.6\linewidth]{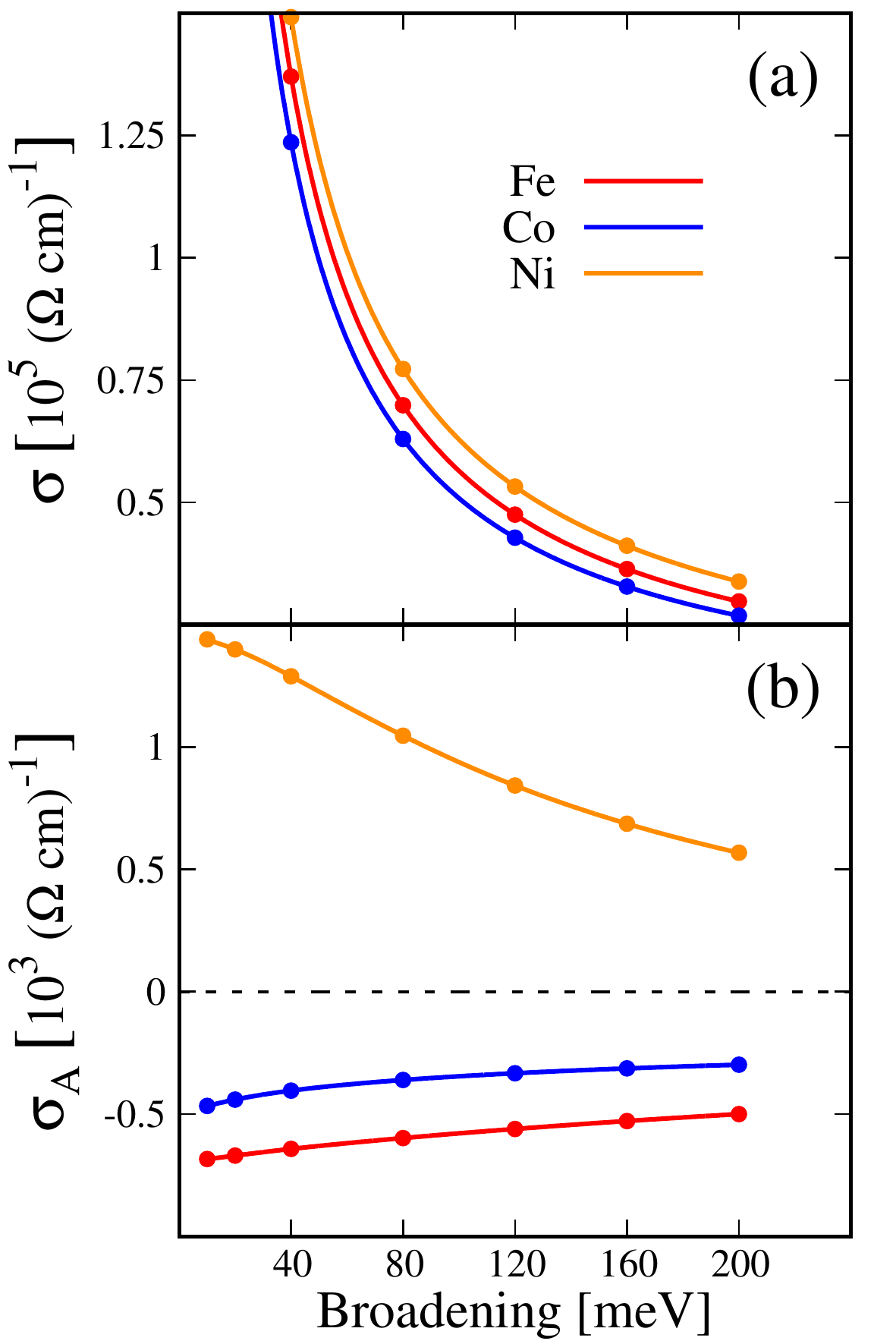}
  \caption{Influence of the lifetime broadening on (a) the longitudinal conductivity and (b) the anomalous Hall conductivity, for Fe, Co, and Ni. For Co only the longitudinal in-plane conductivity is shown,  the out-of-plane conductivity displaying similar broadening dependency.}
  \label{fig:Conductivities-Broadening}
\end{figure}

The anomalous conductivity elements are odd under time-reversal symmetry ($\mathcal{T}$-odd) and require SOC in the calculations. At the Fermi energy, we have $\sigma_A = -0.64$ for Fe, $\sigma_A = -0.40$ for Co and $\sigma_A = 1.29$ for Ni, in units of $10^{3} (\Omega \,\text{cm})^{-1}$. Although there is a noticeable quantitative difference, with a sign change for Ni (which is consistent with previous investigations \cite{Oppeneer2001,Omori2019}), the spectra $\sigma_A(E)$ show strong qualitative similarities. 
In the case of Fe and Co, the negative dip in the spectrum is located around the Fermi energy, giving a negative $\sigma_A(E=0)$, while for Ni there is a positive peak around the Fermi energy, hence the positive value for $\sigma_A(E=0)$.

The lifetime broadening dependency of $\bm{\sigma}$ is shown in Fig.\ \ref{fig:Conductivities-Broadening}. The longitudinal conductivities [Fig.\ \ref{fig:Conductivities-Broadening}(a)] display a $\propto \delta^{-1}$ scaling, as expected, since this component arises mainly from the intraband response of the electronic states around the Fermi energy. 

For $\sigma_A$ [Fig.\ \ref{fig:Conductivities-Broadening}(b)], the broadening dependency is different, as here the interband contribution of Eq.\ (\ref{eq:LinearResponse}) is responsible, as has been reported in previous works \cite{Nagaosa2010}. We note that extrinsic contributions to $\sigma_A$ such as the side jump or skew scattering are not explicitly included in our calculations.

\section{Spin and orbital transverse thermal conductivities}
\label{thermal-conductivities}

\begin{widetext}

\begin{figure*}[bh!]
  \includegraphics[width=0.93\linewidth]{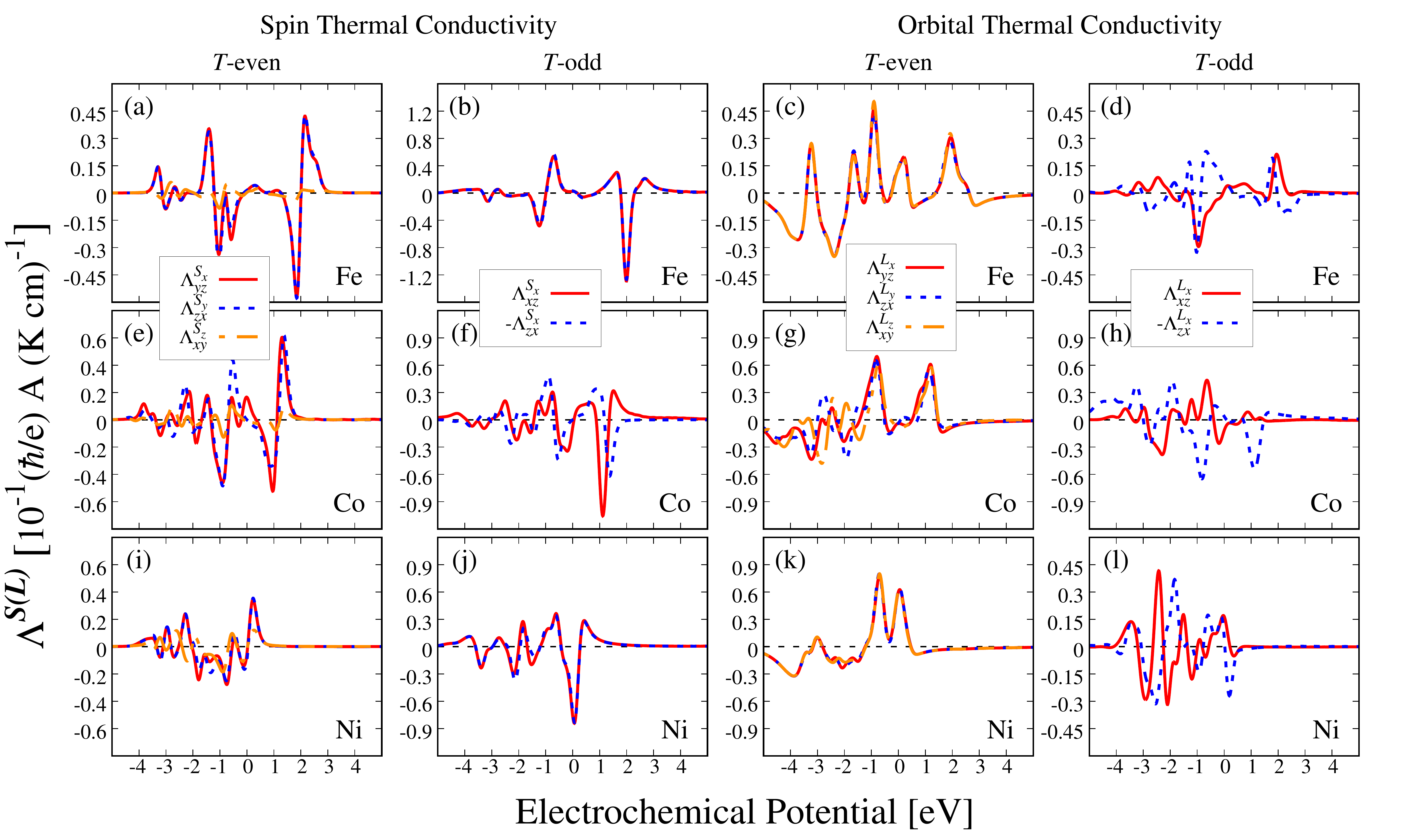}
  \caption{Calculated spin Nernst, magnetic spin Nernst, orbital Nernst and magnetic orbital Nernst effects in Fe, Co, and Ni as a function of the electrochemical potential $E$ at $T =300$ K and for $\hbar \delta = 40 $ meV. }
    \label{fig:Thermal-Conductivities}
\end{figure*}
\end{widetext}

We provide calculated results for the nonzero elements of the $\bm{\Lambda}^{\bm{S}}$ and $\bm{\Lambda}^{\bm{L}}$ tensors, giving the SNE, MSNE, as well as the ONE and MONE, as a function of the electrochemical potential $E$ in Fig.\ \ref{fig:Thermal-Conductivities}. The computed spin and orbital thermal conductivities all display a strong variation with the electrochemical potential. The peak values of the $\mathcal{T}$-odd magnetic and the $\mathcal{T}$-even components are comparably large. 

\FloatBarrier

\newpage

\bibliographystyle{apsrev4-2}
\bibliography{Bibliography}

\begin{thebibliography}{62}%
\makeatletter
\providecommand \@ifxundefined [1]{%
 \@ifx{#1\undefined}
}%
\providecommand \@ifnum [1]{%
 \ifnum #1\expandafter \@firstoftwo
 \else \expandafter \@secondoftwo
 \fi
}%
\providecommand \@ifx [1]{%
 \ifx #1\expandafter \@firstoftwo
 \else \expandafter \@secondoftwo
 \fi
}%
\providecommand \natexlab [1]{#1}%
\providecommand \enquote  [1]{``#1''}%
\providecommand \bibnamefont  [1]{#1}%
\providecommand \bibfnamefont [1]{#1}%
\providecommand \citenamefont [1]{#1}%
\providecommand \href@noop [0]{\@secondoftwo}%
\providecommand \href [0]{\begingroup \@sanitize@url \@href}%
\providecommand \@href[1]{\@@startlink{#1}\@@href}%
\providecommand \@@href[1]{\endgroup#1\@@endlink}%
\providecommand \@sanitize@url [0]{\catcode `\\12\catcode `\$12\catcode
  `\&12\catcode `\#12\catcode `\^12\catcode `\_12\catcode `\%12\relax}%
\providecommand \@@startlink[1]{}%
\providecommand \@@endlink[0]{}%
\providecommand \url  [0]{\begingroup\@sanitize@url \@url }%
\providecommand \@url [1]{\endgroup\@href {#1}{\urlprefix }}%
\providecommand \urlprefix  [0]{URL }%
\providecommand \Eprint [0]{\href }%
\providecommand \doibase [0]{https://doi.org/}%
\providecommand \selectlanguage [0]{\@gobble}%
\providecommand \bibinfo  [0]{\@secondoftwo}%
\providecommand \bibfield  [0]{\@secondoftwo}%
\providecommand \translation [1]{[#1]}%
\providecommand \BibitemOpen [0]{}%
\providecommand \bibitemStop [0]{}%
\providecommand \bibitemNoStop [0]{.\EOS\space}%
\providecommand \EOS [0]{\spacefactor3000\relax}%
\providecommand \BibitemShut  [1]{\csname bibitem#1\endcsname}%
\let\auto@bib@innerbib\@empty
\bibitem [{\citenamefont {Sinova}\ \emph {et~al.}(2015)\citenamefont {Sinova},
  \citenamefont {Valenzuela}, \citenamefont {Wunderlich}, \citenamefont
  {Back},\ and\ \citenamefont {Jungwirth}}]{Sinova2015}%
  \BibitemOpen
  \bibfield  {author} {\bibinfo {author} {\bibfnamefont {J.}~\bibnamefont
  {Sinova}}, \bibinfo {author} {\bibfnamefont {S.~O.}\ \bibnamefont
  {Valenzuela}}, \bibinfo {author} {\bibfnamefont {J.}~\bibnamefont
  {Wunderlich}}, \bibinfo {author} {\bibfnamefont {C.~H.}\ \bibnamefont
  {Back}},\ and\ \bibinfo {author} {\bibfnamefont {T.}~\bibnamefont
  {Jungwirth}},\ }\href {https://doi.org/10.1103/RevModPhys.87.1213} {\bibfield
   {journal} {\bibinfo  {journal} {Rev. Mod. Phys.}\ }\textbf {\bibinfo
  {volume} {87}},\ \bibinfo {pages} {1213} (\bibinfo {year}
  {2015})}\BibitemShut {NoStop}%
\bibitem [{\citenamefont {Hoffmann}(2013)}]{Hoffmann2013}%
  \BibitemOpen
  \bibfield  {author} {\bibinfo {author} {\bibfnamefont {A.}~\bibnamefont
  {Hoffmann}},\ }\href@noop {} {\bibfield  {journal} {\bibinfo  {journal} {IEEE
  Trans. Magn.}\ }\textbf {\bibinfo {volume} {49}},\ \bibinfo {pages} {5172}
  (\bibinfo {year} {2013})}\BibitemShut {NoStop}%
\bibitem [{\citenamefont {Dyakonov}\ and\ \citenamefont
  {Perel}(1971{\natexlab{a}})}]{dyakonovPossibilityOrientingElectron1971}%
  \BibitemOpen
  \bibfield  {author} {\bibinfo {author} {\bibfnamefont {M.~I.}\ \bibnamefont
  {Dyakonov}}\ and\ \bibinfo {author} {\bibfnamefont {V.~I.}\ \bibnamefont
  {Perel}},\ }\href@noop {} {\bibfield  {journal} {\bibinfo  {journal} {JETP
  Lett.}\ }\textbf {\bibinfo {volume} {13}},\ \bibinfo {pages} {657} (\bibinfo
  {year} {1971}{\natexlab{a}})}\BibitemShut {NoStop}%
\bibitem [{\citenamefont {Dyakonov}\ and\ \citenamefont
  {Perel}(1971{\natexlab{b}})}]{dyakonovCurrentinducedSpinOrientation1971}%
  \BibitemOpen
  \bibfield  {author} {\bibinfo {author} {\bibfnamefont {M.~I.}\ \bibnamefont
  {Dyakonov}}\ and\ \bibinfo {author} {\bibfnamefont {V.~I.}\ \bibnamefont
  {Perel}},\ }\href@noop {} {\bibfield  {journal} {\bibinfo  {journal} {Phys.
  Lett. A}\ }\textbf {\bibinfo {volume} {35}},\ \bibinfo {pages} {459}
  (\bibinfo {year} {1971}{\natexlab{b}})}\BibitemShut {NoStop}%
\bibitem [{\citenamefont {Hirsch}(1999)}]{hirschSpinHallEffect1999}%
  \BibitemOpen
  \bibfield  {author} {\bibinfo {author} {\bibfnamefont {J.~E.}\ \bibnamefont
  {Hirsch}},\ }\href@noop {} {\bibfield  {journal} {\bibinfo  {journal} {Phys.
  Rev. Lett.}\ }\textbf {\bibinfo {volume} {83}},\ \bibinfo {pages} {1834}
  (\bibinfo {year} {1999})}\BibitemShut {NoStop}%
\bibitem [{\citenamefont {Zhang}(2000)}]{zhangSpinHallEffect2000}%
  \BibitemOpen
  \bibfield  {author} {\bibinfo {author} {\bibfnamefont {S.}~\bibnamefont
  {Zhang}},\ }\href@noop {} {\bibfield  {journal} {\bibinfo  {journal} {Phys.
  Rev. Lett.}\ }\textbf {\bibinfo {volume} {85}},\ \bibinfo {pages} {393}
  (\bibinfo {year} {2000})}\BibitemShut {NoStop}%
\bibitem [{\citenamefont {Kato}\ \emph
  {et~al.}(2004{\natexlab{a}})\citenamefont {Kato}, \citenamefont {Myers},
  \citenamefont {Gossard},\ and\ \citenamefont
  {Awschalom}}]{katoCoherentSpinManipulation2004}%
  \BibitemOpen
  \bibfield  {author} {\bibinfo {author} {\bibfnamefont {Y.}~\bibnamefont
  {Kato}}, \bibinfo {author} {\bibfnamefont {R.~C.}\ \bibnamefont {Myers}},
  \bibinfo {author} {\bibfnamefont {A.~C.}\ \bibnamefont {Gossard}},\ and\
  \bibinfo {author} {\bibfnamefont {D.~D.}\ \bibnamefont {Awschalom}},\
  }\href@noop {} {\bibfield  {journal} {\bibinfo  {journal} {Nature}\ }\textbf
  {\bibinfo {volume} {427}},\ \bibinfo {pages} {50} (\bibinfo {year}
  {2004}{\natexlab{a}})}\BibitemShut {NoStop}%
\bibitem [{\citenamefont {Kato}\ \emph
  {et~al.}(2004{\natexlab{b}})\citenamefont {Kato}, \citenamefont {Myers},
  \citenamefont {Gossard},\ and\ \citenamefont
  {Awschalom}}]{katoCurrentInducedSpinPolarization2004}%
  \BibitemOpen
  \bibfield  {author} {\bibinfo {author} {\bibfnamefont {Y.~K.}\ \bibnamefont
  {Kato}}, \bibinfo {author} {\bibfnamefont {R.~C.}\ \bibnamefont {Myers}},
  \bibinfo {author} {\bibfnamefont {A.~C.}\ \bibnamefont {Gossard}},\ and\
  \bibinfo {author} {\bibfnamefont {D.~D.}\ \bibnamefont {Awschalom}},\
  }\href@noop {} {\bibfield  {journal} {\bibinfo  {journal} {Phys. Rev. Lett.}\
  }\textbf {\bibinfo {volume} {93}},\ \bibinfo {pages} {176601} (\bibinfo
  {year} {2004}{\natexlab{b}})}\BibitemShut {NoStop}%
\bibitem [{\citenamefont {Stephens}\ \emph {et~al.}(2004)\citenamefont
  {Stephens}, \citenamefont {Berezovsky}, \citenamefont {McGuire},
  \citenamefont {Sham}, \citenamefont {Gossard},\ and\ \citenamefont
  {Awschalom}}]{stephensSpinAccumulationForwardBiased2004}%
  \BibitemOpen
  \bibfield  {author} {\bibinfo {author} {\bibfnamefont {J.}~\bibnamefont
  {Stephens}}, \bibinfo {author} {\bibfnamefont {J.}~\bibnamefont
  {Berezovsky}}, \bibinfo {author} {\bibfnamefont {J.~P.}\ \bibnamefont
  {McGuire}}, \bibinfo {author} {\bibfnamefont {L.~J.}\ \bibnamefont {Sham}},
  \bibinfo {author} {\bibfnamefont {A.~C.}\ \bibnamefont {Gossard}},\ and\
  \bibinfo {author} {\bibfnamefont {D.~D.}\ \bibnamefont {Awschalom}},\
  }\href@noop {} {\bibfield  {journal} {\bibinfo  {journal} {Phys. Rev. Lett.}\
  }\textbf {\bibinfo {volume} {93}},\ \bibinfo {pages} {097602} (\bibinfo
  {year} {2004})}\BibitemShut {NoStop}%
\bibitem [{\citenamefont {Saitoh}\ \emph {et~al.}(2006)\citenamefont {Saitoh},
  \citenamefont {Ueda}, \citenamefont {Miyajima},\ and\ \citenamefont
  {Tatara}}]{saitohConversionSpinCurrent2006}%
  \BibitemOpen
  \bibfield  {author} {\bibinfo {author} {\bibfnamefont {E.}~\bibnamefont
  {Saitoh}}, \bibinfo {author} {\bibfnamefont {M.}~\bibnamefont {Ueda}},
  \bibinfo {author} {\bibfnamefont {H.}~\bibnamefont {Miyajima}},\ and\
  \bibinfo {author} {\bibfnamefont {G.}~\bibnamefont {Tatara}},\ }\href@noop {}
  {\bibfield  {journal} {\bibinfo  {journal} {Appl. Phys. Lett.}\ }\textbf
  {\bibinfo {volume} {88}},\ \bibinfo {pages} {182509} (\bibinfo {year}
  {2006})}\BibitemShut {NoStop}%
\bibitem [{\citenamefont {Kimura}\ \emph {et~al.}(2007)\citenamefont {Kimura},
  \citenamefont {Otani}, \citenamefont {Sato}, \citenamefont {Takahashi},\ and\
  \citenamefont {Maekawa}}]{kimuraRoomTemperatureReversibleSpin2007}%
  \BibitemOpen
  \bibfield  {author} {\bibinfo {author} {\bibfnamefont {T.}~\bibnamefont
  {Kimura}}, \bibinfo {author} {\bibfnamefont {Y.}~\bibnamefont {Otani}},
  \bibinfo {author} {\bibfnamefont {T.}~\bibnamefont {Sato}}, \bibinfo {author}
  {\bibfnamefont {S.}~\bibnamefont {Takahashi}},\ and\ \bibinfo {author}
  {\bibfnamefont {S.}~\bibnamefont {Maekawa}},\ }\href@noop {} {\bibfield
  {journal} {\bibinfo  {journal} {Phys. Rev. Lett.}\ }\textbf {\bibinfo
  {volume} {98}},\ \bibinfo {pages} {156601} (\bibinfo {year}
  {2007})}\BibitemShut {NoStop}%
\bibitem [{\citenamefont {Seki}\ \emph {et~al.}(2008)\citenamefont {Seki},
  \citenamefont {Hasegawa}, \citenamefont {Mitani}, \citenamefont {Takahashi},
  \citenamefont {Imamura}, \citenamefont {Maekawa}, \citenamefont {Nitta},\
  and\ \citenamefont {Takanashi}}]{sekiGiantSpinHall2008}%
  \BibitemOpen
  \bibfield  {author} {\bibinfo {author} {\bibfnamefont {T.}~\bibnamefont
  {Seki}}, \bibinfo {author} {\bibfnamefont {Y.}~\bibnamefont {Hasegawa}},
  \bibinfo {author} {\bibfnamefont {S.}~\bibnamefont {Mitani}}, \bibinfo
  {author} {\bibfnamefont {S.}~\bibnamefont {Takahashi}}, \bibinfo {author}
  {\bibfnamefont {H.}~\bibnamefont {Imamura}}, \bibinfo {author} {\bibfnamefont
  {S.}~\bibnamefont {Maekawa}}, \bibinfo {author} {\bibfnamefont
  {J.}~\bibnamefont {Nitta}},\ and\ \bibinfo {author} {\bibfnamefont
  {K.}~\bibnamefont {Takanashi}},\ }\href@noop {} {\bibfield  {journal}
  {\bibinfo  {journal} {Nature Mater}\ }\textbf {\bibinfo {volume} {7}},\
  \bibinfo {pages} {125} (\bibinfo {year} {2008})}\BibitemShut {NoStop}%
\bibitem [{\citenamefont {Stamm}\ \emph {et~al.}(2017)\citenamefont {Stamm},
  \citenamefont {Murer}, \citenamefont {Berritta}, \citenamefont {Feng},
  \citenamefont {Gabureac}, \citenamefont {Oppeneer},\ and\ \citenamefont
  {Gambardella}}]{Stamm2017}%
  \BibitemOpen
  \bibfield  {author} {\bibinfo {author} {\bibfnamefont {C.}~\bibnamefont
  {Stamm}}, \bibinfo {author} {\bibfnamefont {C.}~\bibnamefont {Murer}},
  \bibinfo {author} {\bibfnamefont {M.}~\bibnamefont {Berritta}}, \bibinfo
  {author} {\bibfnamefont {J.}~\bibnamefont {Feng}}, \bibinfo {author}
  {\bibfnamefont {M.}~\bibnamefont {Gabureac}}, \bibinfo {author}
  {\bibfnamefont {P.~M.}\ \bibnamefont {Oppeneer}},\ and\ \bibinfo {author}
  {\bibfnamefont {P.}~\bibnamefont {Gambardella}},\ }\href
  {https://doi.org/10.1103/PhysRevLett.119.087203} {\bibfield  {journal}
  {\bibinfo  {journal} {Phys. Rev. Lett.}\ }\textbf {\bibinfo {volume} {119}},\
  \bibinfo {pages} {087203} (\bibinfo {year} {2017})}\BibitemShut {NoStop}%
\bibitem [{\citenamefont {{Mihai Miron}}\ \emph {et~al.}(2011)\citenamefont
  {{Mihai Miron}}, \citenamefont {Garello}, \citenamefont {Gaudin},
  \citenamefont {Zermatten}, \citenamefont {Costache}, \citenamefont {Auffret},
  \citenamefont {Bandiera}, \citenamefont {Rodmacq}, \citenamefont {Schuhl},\
  and\ \citenamefont {Gambardella}}]{MihaiMiron2011}%
  \BibitemOpen
  \bibfield  {author} {\bibinfo {author} {\bibfnamefont {I.}~\bibnamefont
  {{Mihai Miron}}}, \bibinfo {author} {\bibfnamefont {K.}~\bibnamefont
  {Garello}}, \bibinfo {author} {\bibfnamefont {G.}~\bibnamefont {Gaudin}},
  \bibinfo {author} {\bibfnamefont {P.~J.}\ \bibnamefont {Zermatten}}, \bibinfo
  {author} {\bibfnamefont {M.~V.}\ \bibnamefont {Costache}}, \bibinfo {author}
  {\bibfnamefont {S.}~\bibnamefont {Auffret}}, \bibinfo {author} {\bibfnamefont
  {S.}~\bibnamefont {Bandiera}}, \bibinfo {author} {\bibfnamefont
  {B.}~\bibnamefont {Rodmacq}}, \bibinfo {author} {\bibfnamefont
  {A.}~\bibnamefont {Schuhl}},\ and\ \bibinfo {author} {\bibfnamefont
  {P.}~\bibnamefont {Gambardella}},\ }\href
  {https://doi.org/10.1038/nature10309} {\bibfield  {journal} {\bibinfo
  {journal} {Nature}\ }\textbf {\bibinfo {volume} {476}},\ \bibinfo {pages}
  {189} (\bibinfo {year} {2011})}\BibitemShut {NoStop}%
\bibitem [{\citenamefont {Liu}\ \emph {et~al.}(2012)\citenamefont {Liu},
  \citenamefont {Lee}, \citenamefont {Gudmundsen}, \citenamefont {Ralph},\ and\
  \citenamefont {Buhrman}}]{liuCurrentInducedSwitchingPerpendicularly2012}%
  \BibitemOpen
  \bibfield  {author} {\bibinfo {author} {\bibfnamefont {L.}~\bibnamefont
  {Liu}}, \bibinfo {author} {\bibfnamefont {O.~J.}\ \bibnamefont {Lee}},
  \bibinfo {author} {\bibfnamefont {T.~J.}\ \bibnamefont {Gudmundsen}},
  \bibinfo {author} {\bibfnamefont {D.~C.}\ \bibnamefont {Ralph}},\ and\
  \bibinfo {author} {\bibfnamefont {R.~A.}\ \bibnamefont {Buhrman}},\
  }\href@noop {} {\bibfield  {journal} {\bibinfo  {journal} {Phys. Rev. Lett.}\
  }\textbf {\bibinfo {volume} {109}},\ \bibinfo {pages} {096602} (\bibinfo
  {year} {2012})}\BibitemShut {NoStop}%
\bibitem [{\citenamefont {Jamali}\ \emph {et~al.}(2013)\citenamefont {Jamali},
  \citenamefont {Narayanapillai}, \citenamefont {Qiu}, \citenamefont {Loong},
  \citenamefont {Manchon},\ and\ \citenamefont
  {Yang}}]{jamaliSpinOrbitTorquesCo2013}%
  \BibitemOpen
  \bibfield  {author} {\bibinfo {author} {\bibfnamefont {M.}~\bibnamefont
  {Jamali}}, \bibinfo {author} {\bibfnamefont {K.}~\bibnamefont
  {Narayanapillai}}, \bibinfo {author} {\bibfnamefont {X.}~\bibnamefont {Qiu}},
  \bibinfo {author} {\bibfnamefont {L.~M.}\ \bibnamefont {Loong}}, \bibinfo
  {author} {\bibfnamefont {A.}~\bibnamefont {Manchon}},\ and\ \bibinfo {author}
  {\bibfnamefont {H.}~\bibnamefont {Yang}},\ }\href@noop {} {\bibfield
  {journal} {\bibinfo  {journal} {Phys. Rev. Lett.}\ }\textbf {\bibinfo
  {volume} {111}},\ \bibinfo {pages} {246602} (\bibinfo {year}
  {2013})}\BibitemShut {NoStop}%
\bibitem [{\citenamefont {Fan}\ \emph {et~al.}(2014)\citenamefont {Fan},
  \citenamefont {Upadhyaya}, \citenamefont {Kou}, \citenamefont {Lang},
  \citenamefont {Takei}, \citenamefont {Wang}, \citenamefont {Tang},
  \citenamefont {He}, \citenamefont {Chang}, \citenamefont {Montazeri},
  \citenamefont {Yu}, \citenamefont {Jiang}, \citenamefont {Nie}, \citenamefont
  {Schwartz}, \citenamefont {Tserkovnyak},\ and\ \citenamefont
  {Wang}}]{fanMagnetizationSwitchingGiant2014}%
  \BibitemOpen
  \bibfield  {author} {\bibinfo {author} {\bibfnamefont {Y.}~\bibnamefont
  {Fan}}, \bibinfo {author} {\bibfnamefont {P.}~\bibnamefont {Upadhyaya}},
  \bibinfo {author} {\bibfnamefont {X.}~\bibnamefont {Kou}}, \bibinfo {author}
  {\bibfnamefont {M.}~\bibnamefont {Lang}}, \bibinfo {author} {\bibfnamefont
  {S.}~\bibnamefont {Takei}}, \bibinfo {author} {\bibfnamefont
  {Z.}~\bibnamefont {Wang}}, \bibinfo {author} {\bibfnamefont {J.}~\bibnamefont
  {Tang}}, \bibinfo {author} {\bibfnamefont {L.}~\bibnamefont {He}}, \bibinfo
  {author} {\bibfnamefont {L.-T.}\ \bibnamefont {Chang}}, \bibinfo {author}
  {\bibfnamefont {M.}~\bibnamefont {Montazeri}}, \bibinfo {author}
  {\bibfnamefont {G.}~\bibnamefont {Yu}}, \bibinfo {author} {\bibfnamefont
  {W.}~\bibnamefont {Jiang}}, \bibinfo {author} {\bibfnamefont
  {T.}~\bibnamefont {Nie}}, \bibinfo {author} {\bibfnamefont {R.~N.}\
  \bibnamefont {Schwartz}}, \bibinfo {author} {\bibfnamefont {Y.}~\bibnamefont
  {Tserkovnyak}},\ and\ \bibinfo {author} {\bibfnamefont {K.~L.}\ \bibnamefont
  {Wang}},\ }\href@noop {} {\bibfield  {journal} {\bibinfo  {journal} {Nature
  Mater}\ }\textbf {\bibinfo {volume} {13}},\ \bibinfo {pages} {699} (\bibinfo
  {year} {2014})}\BibitemShut {NoStop}%
\bibitem [{\citenamefont {Garello}\ \emph {et~al.}(2014)\citenamefont
  {Garello}, \citenamefont {Avci}, \citenamefont {Miron}, \citenamefont
  {Baumgartner}, \citenamefont {Ghosh}, \citenamefont {Auffret}, \citenamefont
  {Boulle}, \citenamefont {Gaudin},\ and\ \citenamefont
  {Gambardella}}]{garelloUltrafastMagnetizationSwitching2014}%
  \BibitemOpen
  \bibfield  {author} {\bibinfo {author} {\bibfnamefont {K.}~\bibnamefont
  {Garello}}, \bibinfo {author} {\bibfnamefont {C.~O.}\ \bibnamefont {Avci}},
  \bibinfo {author} {\bibfnamefont {I.~M.}\ \bibnamefont {Miron}}, \bibinfo
  {author} {\bibfnamefont {M.}~\bibnamefont {Baumgartner}}, \bibinfo {author}
  {\bibfnamefont {A.}~\bibnamefont {Ghosh}}, \bibinfo {author} {\bibfnamefont
  {S.}~\bibnamefont {Auffret}}, \bibinfo {author} {\bibfnamefont
  {O.}~\bibnamefont {Boulle}}, \bibinfo {author} {\bibfnamefont
  {G.}~\bibnamefont {Gaudin}},\ and\ \bibinfo {author} {\bibfnamefont
  {P.}~\bibnamefont {Gambardella}},\ }\href@noop {} {\bibfield  {journal}
  {\bibinfo  {journal} {Appl. Phys. Lett.}\ }\textbf {\bibinfo {volume}
  {105}},\ \bibinfo {pages} {212402} (\bibinfo {year} {2014})}\BibitemShut
  {NoStop}%
\bibitem [{\citenamefont {Hao}\ and\ \citenamefont
  {Xiao}(2015)}]{haoGiantSpinHall2015}%
  \BibitemOpen
  \bibfield  {author} {\bibinfo {author} {\bibfnamefont {Q.}~\bibnamefont
  {Hao}}\ and\ \bibinfo {author} {\bibfnamefont {G.}~\bibnamefont {Xiao}},\
  }\href@noop {} {\bibfield  {journal} {\bibinfo  {journal} {Phys. Rev.
  Applied}\ }\textbf {\bibinfo {volume} {3}},\ \bibinfo {pages} {034009}
  (\bibinfo {year} {2015})}\BibitemShut {NoStop}%
\bibitem [{\citenamefont {Lee}\ \emph {et~al.}(2014)\citenamefont {Lee},
  \citenamefont {Lee}, \citenamefont {Min},\ and\ \citenamefont
  {Lee}}]{leeThermallyActivatedSwitching2014}%
  \BibitemOpen
  \bibfield  {author} {\bibinfo {author} {\bibfnamefont {K.-S.}\ \bibnamefont
  {Lee}}, \bibinfo {author} {\bibfnamefont {S.-W.}\ \bibnamefont {Lee}},
  \bibinfo {author} {\bibfnamefont {B.-C.}\ \bibnamefont {Min}},\ and\ \bibinfo
  {author} {\bibfnamefont {K.-J.}\ \bibnamefont {Lee}},\ }\href@noop {}
  {\bibfield  {journal} {\bibinfo  {journal} {Appl. Phys. Lett.}\ }\textbf
  {\bibinfo {volume} {104}},\ \bibinfo {pages} {072413} (\bibinfo {year}
  {2014})}\BibitemShut {NoStop}%
\bibitem [{\citenamefont {Karplus}\ and\ \citenamefont
  {Luttinger}(1954)}]{karplusHallEffectFerromagnetics1954}%
  \BibitemOpen
  \bibfield  {author} {\bibinfo {author} {\bibfnamefont {R.}~\bibnamefont
  {Karplus}}\ and\ \bibinfo {author} {\bibfnamefont {J.~M.}\ \bibnamefont
  {Luttinger}},\ }\href@noop {} {\bibfield  {journal} {\bibinfo  {journal}
  {Phys. Rev.}\ }\textbf {\bibinfo {volume} {95}},\ \bibinfo {pages} {1154}
  (\bibinfo {year} {1954})}\BibitemShut {NoStop}%
\bibitem [{\citenamefont
  {Murakami}(2003)}]{murakamiDissipationlessQuantumSpin2003}%
  \BibitemOpen
  \bibfield  {author} {\bibinfo {author} {\bibfnamefont {S.}~\bibnamefont
  {Murakami}},\ }\href@noop {} {\bibfield  {journal} {\bibinfo  {journal}
  {Science}\ }\textbf {\bibinfo {volume} {301}},\ \bibinfo {pages} {1348}
  (\bibinfo {year} {2003})}\BibitemShut {NoStop}%
\bibitem [{\citenamefont {Sinova}\ \emph {et~al.}(2004)\citenamefont {Sinova},
  \citenamefont {Culcer}, \citenamefont {Niu}, \citenamefont {Sinitsyn},
  \citenamefont {Jungwirth},\ and\ \citenamefont
  {MacDonald}}]{sinovaUniversalIntrinsicSpin2004}%
  \BibitemOpen
  \bibfield  {author} {\bibinfo {author} {\bibfnamefont {J.}~\bibnamefont
  {Sinova}}, \bibinfo {author} {\bibfnamefont {D.}~\bibnamefont {Culcer}},
  \bibinfo {author} {\bibfnamefont {Q.}~\bibnamefont {Niu}}, \bibinfo {author}
  {\bibfnamefont {N.~A.}\ \bibnamefont {Sinitsyn}}, \bibinfo {author}
  {\bibfnamefont {T.}~\bibnamefont {Jungwirth}},\ and\ \bibinfo {author}
  {\bibfnamefont {A.~H.}\ \bibnamefont {MacDonald}},\ }\href@noop {} {\bibfield
   {journal} {\bibinfo  {journal} {Phys. Rev. Lett.}\ }\textbf {\bibinfo
  {volume} {92}},\ \bibinfo {pages} {126603} (\bibinfo {year}
  {2004})}\BibitemShut {NoStop}%
\bibitem [{\citenamefont {Tanaka}\ \emph
  {et~al.}(2008{\natexlab{a}})\citenamefont {Tanaka}, \citenamefont {Kontani},
  \citenamefont {Naito}, \citenamefont {Naito}, \citenamefont {Hirashima},
  \citenamefont {Yamada},\ and\ \citenamefont
  {Inoue}}]{tanakaIntrinsicSpinHall2008}%
  \BibitemOpen
  \bibfield  {author} {\bibinfo {author} {\bibfnamefont {T.}~\bibnamefont
  {Tanaka}}, \bibinfo {author} {\bibfnamefont {H.}~\bibnamefont {Kontani}},
  \bibinfo {author} {\bibfnamefont {M.}~\bibnamefont {Naito}}, \bibinfo
  {author} {\bibfnamefont {T.}~\bibnamefont {Naito}}, \bibinfo {author}
  {\bibfnamefont {D.~S.}\ \bibnamefont {Hirashima}}, \bibinfo {author}
  {\bibfnamefont {K.}~\bibnamefont {Yamada}},\ and\ \bibinfo {author}
  {\bibfnamefont {J.}~\bibnamefont {Inoue}},\ }\href@noop {} {\bibfield
  {journal} {\bibinfo  {journal} {Phys. Rev. B}\ }\textbf {\bibinfo {volume}
  {77}},\ \bibinfo {pages} {165117} (\bibinfo {year}
  {2008}{\natexlab{a}})}\BibitemShut {NoStop}%
\bibitem [{\citenamefont {Berger}(1970)}]{bergerSideJumpMechanismHall1970}%
  \BibitemOpen
  \bibfield  {author} {\bibinfo {author} {\bibfnamefont {L.}~\bibnamefont
  {Berger}},\ }\href@noop {} {\bibfield  {journal} {\bibinfo  {journal} {Phys.
  Rev. B}\ }\textbf {\bibinfo {volume} {2}},\ \bibinfo {pages} {4559} (\bibinfo
  {year} {1970})}\BibitemShut {NoStop}%
\bibitem [{\citenamefont {Smit}(1958)}]{smitSpontaneousHallEffect1958}%
  \BibitemOpen
  \bibfield  {author} {\bibinfo {author} {\bibfnamefont {J.}~\bibnamefont
  {Smit}},\ }\href@noop {} {\bibfield  {journal} {\bibinfo  {journal}
  {Physica}\ }\textbf {\bibinfo {volume} {24}},\ \bibinfo {pages} {39}
  (\bibinfo {year} {1958})}\BibitemShut {NoStop}%
\bibitem [{\citenamefont {Kontani}\ \emph {et~al.}(2007)\citenamefont
  {Kontani}, \citenamefont {Naito}, \citenamefont {Hirashima}, \citenamefont
  {Yamada},\ and\ \citenamefont {Inoue}}]{kontaniStudyIntrinsicSpin2007}%
  \BibitemOpen
  \bibfield  {author} {\bibinfo {author} {\bibfnamefont {H.}~\bibnamefont
  {Kontani}}, \bibinfo {author} {\bibfnamefont {M.}~\bibnamefont {Naito}},
  \bibinfo {author} {\bibfnamefont {D.~S.}\ \bibnamefont {Hirashima}}, \bibinfo
  {author} {\bibfnamefont {K.}~\bibnamefont {Yamada}},\ and\ \bibinfo {author}
  {\bibfnamefont {J.-I.}\ \bibnamefont {Inoue}},\ }\href@noop {} {\bibfield
  {journal} {\bibinfo  {journal} {J. Phys. Soc. Jpn.}\ }\textbf {\bibinfo
  {volume} {76}},\ \bibinfo {pages} {103702} (\bibinfo {year}
  {2007})}\BibitemShut {NoStop}%
\bibitem [{\citenamefont {Kontani}\ \emph {et~al.}(2009)\citenamefont
  {Kontani}, \citenamefont {Tanaka}, \citenamefont {Hirashima}, \citenamefont
  {Yamada},\ and\ \citenamefont {Inoue}}]{kontaniGiantOrbitalHall2009}%
  \BibitemOpen
  \bibfield  {author} {\bibinfo {author} {\bibfnamefont {H.}~\bibnamefont
  {Kontani}}, \bibinfo {author} {\bibfnamefont {T.}~\bibnamefont {Tanaka}},
  \bibinfo {author} {\bibfnamefont {D.~S.}\ \bibnamefont {Hirashima}}, \bibinfo
  {author} {\bibfnamefont {K.}~\bibnamefont {Yamada}},\ and\ \bibinfo {author}
  {\bibfnamefont {J.}~\bibnamefont {Inoue}},\ }\href@noop {} {\bibfield
  {journal} {\bibinfo  {journal} {Phys. Rev. Lett.}\ }\textbf {\bibinfo
  {volume} {102}},\ \bibinfo {pages} {016601} (\bibinfo {year}
  {2009})}\BibitemShut {NoStop}%
\bibitem [{\citenamefont {Kontani}\ \emph {et~al.}(2008)\citenamefont
  {Kontani}, \citenamefont {Tanaka}, \citenamefont {Hirashima}, \citenamefont
  {Yamada},\ and\ \citenamefont {Inoue}}]{kontaniGiantIntrinsicSpin2008}%
  \BibitemOpen
  \bibfield  {author} {\bibinfo {author} {\bibfnamefont {H.}~\bibnamefont
  {Kontani}}, \bibinfo {author} {\bibfnamefont {T.}~\bibnamefont {Tanaka}},
  \bibinfo {author} {\bibfnamefont {D.~S.}\ \bibnamefont {Hirashima}}, \bibinfo
  {author} {\bibfnamefont {K.}~\bibnamefont {Yamada}},\ and\ \bibinfo {author}
  {\bibfnamefont {J.}~\bibnamefont {Inoue}},\ }\href@noop {} {\bibfield
  {journal} {\bibinfo  {journal} {Phys. Rev. Lett.}\ }\textbf {\bibinfo
  {volume} {100}},\ \bibinfo {pages} {096601} (\bibinfo {year}
  {2008})}\BibitemShut {NoStop}%
\bibitem [{\citenamefont {Tanaka}\ and\ \citenamefont
  {Kontani}(2010)}]{tanakaIntrinsicSpinOrbital2010}%
  \BibitemOpen
  \bibfield  {author} {\bibinfo {author} {\bibfnamefont {T.}~\bibnamefont
  {Tanaka}}\ and\ \bibinfo {author} {\bibfnamefont {H.}~\bibnamefont
  {Kontani}},\ }\href@noop {} {\bibfield  {journal} {\bibinfo  {journal} {Phys.
  Rev. B}\ }\textbf {\bibinfo {volume} {81}},\ \bibinfo {pages} {224401}
  (\bibinfo {year} {2010})}\BibitemShut {NoStop}%
\bibitem [{\citenamefont {Go}\ \emph {et~al.}(2018)\citenamefont {Go},
  \citenamefont {Jo}, \citenamefont {Kim},\ and\ \citenamefont
  {Lee}}]{goIntrinsicSpinOrbital2018}%
  \BibitemOpen
  \bibfield  {author} {\bibinfo {author} {\bibfnamefont {D.}~\bibnamefont
  {Go}}, \bibinfo {author} {\bibfnamefont {D.}~\bibnamefont {Jo}}, \bibinfo
  {author} {\bibfnamefont {C.}~\bibnamefont {Kim}},\ and\ \bibinfo {author}
  {\bibfnamefont {H.-W.}\ \bibnamefont {Lee}},\ }\href@noop {} {\bibfield
  {journal} {\bibinfo  {journal} {Phys. Rev. Lett.}\ }\textbf {\bibinfo
  {volume} {121}},\ \bibinfo {pages} {086602} (\bibinfo {year}
  {2018})}\BibitemShut {NoStop}%
\bibitem [{\citenamefont {Jo}\ \emph {et~al.}(2018)\citenamefont {Jo},
  \citenamefont {Go},\ and\ \citenamefont
  {Lee}}]{joGiganticIntrinsicOrbital2018}%
  \BibitemOpen
  \bibfield  {author} {\bibinfo {author} {\bibfnamefont {D.}~\bibnamefont
  {Jo}}, \bibinfo {author} {\bibfnamefont {D.}~\bibnamefont {Go}},\ and\
  \bibinfo {author} {\bibfnamefont {H.-W.}\ \bibnamefont {Lee}},\ }\href@noop
  {} {\bibfield  {journal} {\bibinfo  {journal} {Phys. Rev. B}\ }\textbf
  {\bibinfo {volume} {98}},\ \bibinfo {pages} {214405} (\bibinfo {year}
  {2018})}\BibitemShut {NoStop}%
\bibitem [{\citenamefont {Humphries}\ \emph {et~al.}(2017)\citenamefont
  {Humphries}, \citenamefont {Wang}, \citenamefont {Edwards}, \citenamefont
  {Allen}, \citenamefont {Shaw}, \citenamefont {Nembach}, \citenamefont {Xiao},
  \citenamefont {Silva},\ and\ \citenamefont
  {Fan}}]{humphriesObservationSpinorbitEffects2017}%
  \BibitemOpen
  \bibfield  {author} {\bibinfo {author} {\bibfnamefont {A.~M.}\ \bibnamefont
  {Humphries}}, \bibinfo {author} {\bibfnamefont {T.}~\bibnamefont {Wang}},
  \bibinfo {author} {\bibfnamefont {E.~R.~J.}\ \bibnamefont {Edwards}},
  \bibinfo {author} {\bibfnamefont {S.~R.}\ \bibnamefont {Allen}}, \bibinfo
  {author} {\bibfnamefont {J.~M.}\ \bibnamefont {Shaw}}, \bibinfo {author}
  {\bibfnamefont {H.~T.}\ \bibnamefont {Nembach}}, \bibinfo {author}
  {\bibfnamefont {J.~Q.}\ \bibnamefont {Xiao}}, \bibinfo {author}
  {\bibfnamefont {T.~J.}\ \bibnamefont {Silva}},\ and\ \bibinfo {author}
  {\bibfnamefont {X.}~\bibnamefont {Fan}},\ }\href@noop {} {\bibfield
  {journal} {\bibinfo  {journal} {Nat Commun}\ }\textbf {\bibinfo {volume}
  {8}},\ \bibinfo {pages} {911} (\bibinfo {year} {2017})}\BibitemShut {NoStop}%
\bibitem [{\citenamefont {Wang}\ \emph {et~al.}(2019)\citenamefont {Wang},
  \citenamefont {Wang}, \citenamefont {Amin}, \citenamefont {Wang},
  \citenamefont {Radhakrishnan}, \citenamefont {Davidson}, \citenamefont
  {Allen}, \citenamefont {Silva}, \citenamefont {Ohldag}, \citenamefont
  {Balzar}, \citenamefont {Zink}, \citenamefont {Haney}, \citenamefont {Xiao},
  \citenamefont {Cahill}, \citenamefont {Lorenz},\ and\ \citenamefont
  {Fan}}]{Wang2019}%
  \BibitemOpen
  \bibfield  {author} {\bibinfo {author} {\bibfnamefont {W.}~\bibnamefont
  {Wang}}, \bibinfo {author} {\bibfnamefont {T.}~\bibnamefont {Wang}}, \bibinfo
  {author} {\bibfnamefont {V.~P.}\ \bibnamefont {Amin}}, \bibinfo {author}
  {\bibfnamefont {Y.}~\bibnamefont {Wang}}, \bibinfo {author} {\bibfnamefont
  {A.}~\bibnamefont {Radhakrishnan}}, \bibinfo {author} {\bibfnamefont
  {A.}~\bibnamefont {Davidson}}, \bibinfo {author} {\bibfnamefont {S.~R.}\
  \bibnamefont {Allen}}, \bibinfo {author} {\bibfnamefont {T.~J.}\ \bibnamefont
  {Silva}}, \bibinfo {author} {\bibfnamefont {H.}~\bibnamefont {Ohldag}},
  \bibinfo {author} {\bibfnamefont {D.}~\bibnamefont {Balzar}}, \bibinfo
  {author} {\bibfnamefont {B.~L.}\ \bibnamefont {Zink}}, \bibinfo {author}
  {\bibfnamefont {P.~M.}\ \bibnamefont {Haney}}, \bibinfo {author}
  {\bibfnamefont {J.~Q.}\ \bibnamefont {Xiao}}, \bibinfo {author}
  {\bibfnamefont {D.~G.}\ \bibnamefont {Cahill}}, \bibinfo {author}
  {\bibfnamefont {V.~O.}\ \bibnamefont {Lorenz}},\ and\ \bibinfo {author}
  {\bibfnamefont {X.}~\bibnamefont {Fan}},\ }\href@noop {} {\bibfield
  {journal} {\bibinfo  {journal} {Nature Nanotechn.}\ }\textbf {\bibinfo
  {volume} {14}},\ \bibinfo {pages} {819} (\bibinfo {year} {2019})}\BibitemShut
  {NoStop}%
\bibitem [{\citenamefont {Davidson}\ \emph {et~al.}(2020)\citenamefont
  {Davidson}, \citenamefont {Amin}, \citenamefont {Aljuaid}, \citenamefont
  {Haney},\ and\ \citenamefont
  {Fan}}]{davidsonPerspectivesElectricallyGenerated2020}%
  \BibitemOpen
  \bibfield  {author} {\bibinfo {author} {\bibfnamefont {A.}~\bibnamefont
  {Davidson}}, \bibinfo {author} {\bibfnamefont {V.~P.}\ \bibnamefont {Amin}},
  \bibinfo {author} {\bibfnamefont {W.~S.}\ \bibnamefont {Aljuaid}}, \bibinfo
  {author} {\bibfnamefont {P.~M.}\ \bibnamefont {Haney}},\ and\ \bibinfo
  {author} {\bibfnamefont {X.}~\bibnamefont {Fan}},\ }\href@noop {} {\bibfield
  {journal} {\bibinfo  {journal} {Phys. Lett. A}\ }\textbf {\bibinfo {volume}
  {384}},\ \bibinfo {pages} {126228} (\bibinfo {year} {2020})}\BibitemShut
  {NoStop}%
\bibitem [{\citenamefont {{\v Z}elezn{\'y}}\ \emph {et~al.}(2017)\citenamefont
  {{\v Z}elezn{\'y}}, \citenamefont {Zhang}, \citenamefont {Felser},\ and\
  \citenamefont {Yan}}]{zeleznySpinPolarizedCurrentNoncollinear2017}%
  \BibitemOpen
  \bibfield  {author} {\bibinfo {author} {\bibfnamefont {J.}~\bibnamefont {{\v
  Z}elezn{\'y}}}, \bibinfo {author} {\bibfnamefont {Y.}~\bibnamefont {Zhang}},
  \bibinfo {author} {\bibfnamefont {C.}~\bibnamefont {Felser}},\ and\ \bibinfo
  {author} {\bibfnamefont {B.}~\bibnamefont {Yan}},\ }\href@noop {} {\bibfield
  {journal} {\bibinfo  {journal} {Phys. Rev. Lett.}\ }\textbf {\bibinfo
  {volume} {119}},\ \bibinfo {pages} {187204} (\bibinfo {year}
  {2017})}\BibitemShut {NoStop}%
\bibitem [{\citenamefont {{\v Z}elezn{\'y}}\ \emph {et~al.}(2018)\citenamefont
  {{\v Z}elezn{\'y}}, \citenamefont {Wadley}, \citenamefont {Olejn{\'i}k},
  \citenamefont {Hoffmann},\ and\ \citenamefont
  {Ohno}}]{zeleznySpinTransportSpin2018}%
  \BibitemOpen
  \bibfield  {author} {\bibinfo {author} {\bibfnamefont {J.}~\bibnamefont {{\v
  Z}elezn{\'y}}}, \bibinfo {author} {\bibfnamefont {P.}~\bibnamefont {Wadley}},
  \bibinfo {author} {\bibfnamefont {K.}~\bibnamefont {Olejn{\'i}k}}, \bibinfo
  {author} {\bibfnamefont {A.}~\bibnamefont {Hoffmann}},\ and\ \bibinfo
  {author} {\bibfnamefont {H.}~\bibnamefont {Ohno}},\ }\href@noop {} {\bibfield
   {journal} {\bibinfo  {journal} {Nature Phys.}\ }\textbf {\bibinfo {volume}
  {14}},\ \bibinfo {pages} {220} (\bibinfo {year} {2018})}\BibitemShut
  {NoStop}%
\bibitem [{\citenamefont {Kimata}\ \emph {et~al.}(2019)\citenamefont {Kimata},
  \citenamefont {Chen}, \citenamefont {Kondou}, \citenamefont {Sugimoto},
  \citenamefont {Muduli}, \citenamefont {Ikhlas}, \citenamefont {Omori},
  \citenamefont {Tomita}, \citenamefont {MacDonald}, \citenamefont
  {Nakatsuji},\ and\ \citenamefont
  {Otani}}]{kimataMagneticMagneticInverse2019}%
  \BibitemOpen
  \bibfield  {author} {\bibinfo {author} {\bibfnamefont {M.}~\bibnamefont
  {Kimata}}, \bibinfo {author} {\bibfnamefont {H.}~\bibnamefont {Chen}},
  \bibinfo {author} {\bibfnamefont {K.}~\bibnamefont {Kondou}}, \bibinfo
  {author} {\bibfnamefont {S.}~\bibnamefont {Sugimoto}}, \bibinfo {author}
  {\bibfnamefont {P.~K.}\ \bibnamefont {Muduli}}, \bibinfo {author}
  {\bibfnamefont {M.}~\bibnamefont {Ikhlas}}, \bibinfo {author} {\bibfnamefont
  {Y.}~\bibnamefont {Omori}}, \bibinfo {author} {\bibfnamefont
  {T.}~\bibnamefont {Tomita}}, \bibinfo {author} {\bibfnamefont {A.~H.}\
  \bibnamefont {MacDonald}}, \bibinfo {author} {\bibfnamefont {S.}~\bibnamefont
  {Nakatsuji}},\ and\ \bibinfo {author} {\bibfnamefont {Y.}~\bibnamefont
  {Otani}},\ }\href@noop {} {\bibfield  {journal} {\bibinfo  {journal}
  {Nature}\ }\textbf {\bibinfo {volume} {565}},\ \bibinfo {pages} {627}
  (\bibinfo {year} {2019})}\BibitemShut {NoStop}%
\bibitem [{\citenamefont {Mook}\ \emph {et~al.}(2020)\citenamefont {Mook},
  \citenamefont {Neumann}, \citenamefont {Johansson}, \citenamefont {Henk},\
  and\ \citenamefont {Mertig}}]{mookOriginMagneticSpin2020}%
  \BibitemOpen
  \bibfield  {author} {\bibinfo {author} {\bibfnamefont {A.}~\bibnamefont
  {Mook}}, \bibinfo {author} {\bibfnamefont {R.~R.}\ \bibnamefont {Neumann}},
  \bibinfo {author} {\bibfnamefont {A.}~\bibnamefont {Johansson}}, \bibinfo
  {author} {\bibfnamefont {J.}~\bibnamefont {Henk}},\ and\ \bibinfo {author}
  {\bibfnamefont {I.}~\bibnamefont {Mertig}},\ }\href@noop {} {\bibfield
  {journal} {\bibinfo  {journal} {Phys. Rev. Research}\ }\textbf {\bibinfo
  {volume} {2}},\ \bibinfo {pages} {023065} (\bibinfo {year}
  {2020})}\BibitemShut {NoStop}%
\bibitem [{\citenamefont {Salemi}\ \emph {et~al.}(2021)\citenamefont {Salemi},
  \citenamefont {Berritta},\ and\ \citenamefont {Oppeneer}}]{Salemi2021}%
  \BibitemOpen
  \bibfield  {author} {\bibinfo {author} {\bibfnamefont {L.}~\bibnamefont
  {Salemi}}, \bibinfo {author} {\bibfnamefont {M.}~\bibnamefont {Berritta}},\
  and\ \bibinfo {author} {\bibfnamefont {P.~M.}\ \bibnamefont {Oppeneer}},\
  }\href {https://doi.org/10.1103/PhysRevMaterials.5.074407} {\bibfield
  {journal} {\bibinfo  {journal} {Phys. Rev. Materials}\ }\textbf {\bibinfo
  {volume} {5}},\ \bibinfo {pages} {074407} (\bibinfo {year}
  {2021})}\BibitemShut {NoStop}%
\bibitem [{\citenamefont {Seemann}\ \emph {et~al.}(2015)\citenamefont
  {Seemann}, \citenamefont {K{\"o}dderitzsch}, \citenamefont {Wimmer},\ and\
  \citenamefont {Ebert}}]{seemannSymmetryimposedShapeLinear2015}%
  \BibitemOpen
  \bibfield  {author} {\bibinfo {author} {\bibfnamefont {M.}~\bibnamefont
  {Seemann}}, \bibinfo {author} {\bibfnamefont {D.}~\bibnamefont
  {K{\"o}dderitzsch}}, \bibinfo {author} {\bibfnamefont {S.}~\bibnamefont
  {Wimmer}},\ and\ \bibinfo {author} {\bibfnamefont {H.}~\bibnamefont
  {Ebert}},\ }\href@noop {} {\bibfield  {journal} {\bibinfo  {journal} {Phys.
  Rev. B}\ }\textbf {\bibinfo {volume} {92}},\ \bibinfo {pages} {155138}
  (\bibinfo {year} {2015})}\BibitemShut {NoStop}%
\bibitem [{\citenamefont {Wang}(2021)}]{Wang2021}%
  \BibitemOpen
  \bibfield  {author} {\bibinfo {author} {\bibfnamefont {X.~R.}\ \bibnamefont
  {Wang}},\ }\href {https://doi.org/10.1038/s42005-021-00557-9} {\bibfield
  {journal} {\bibinfo  {journal} {Commun. Phys.}\ }\textbf {\bibinfo {volume}
  {4}},\ \bibinfo {pages} {55} (\bibinfo {year} {2021})}\BibitemShut {NoStop}%
\bibitem [{\citenamefont {Meyer}\ \emph {et~al.}(2017)\citenamefont {Meyer},
  \citenamefont {Chen}, \citenamefont {Wimmer}, \citenamefont {Althammer},
  \citenamefont {Wimmer}, \citenamefont {Schlitz}, \citenamefont {Geprags},
  \citenamefont {Huebl}, \citenamefont {Kodderitzsch}, \citenamefont {Ebert},
  \citenamefont {Bauer}, \citenamefont {Gross},\ and\ \citenamefont
  {Goennenwein}}]{Meyer2017}%
  \BibitemOpen
  \bibfield  {author} {\bibinfo {author} {\bibfnamefont {S.}~\bibnamefont
  {Meyer}}, \bibinfo {author} {\bibfnamefont {Y.~T.}\ \bibnamefont {Chen}},
  \bibinfo {author} {\bibfnamefont {S.}~\bibnamefont {Wimmer}}, \bibinfo
  {author} {\bibfnamefont {M.}~\bibnamefont {Althammer}}, \bibinfo {author}
  {\bibfnamefont {T.}~\bibnamefont {Wimmer}}, \bibinfo {author} {\bibfnamefont
  {R.}~\bibnamefont {Schlitz}}, \bibinfo {author} {\bibfnamefont
  {S.}~\bibnamefont {Geprags}}, \bibinfo {author} {\bibfnamefont
  {H.}~\bibnamefont {Huebl}}, \bibinfo {author} {\bibfnamefont
  {D.}~\bibnamefont {Kodderitzsch}}, \bibinfo {author} {\bibfnamefont
  {H.}~\bibnamefont {Ebert}}, \bibinfo {author} {\bibfnamefont {G.~E.~W.}\
  \bibnamefont {Bauer}}, \bibinfo {author} {\bibfnamefont {R.}~\bibnamefont
  {Gross}},\ and\ \bibinfo {author} {\bibfnamefont {S.~T.~B.}\ \bibnamefont
  {Goennenwein}},\ }\href {https://doi.org/10.1038/NMAT4964} {\bibfield
  {journal} {\bibinfo  {journal} {Nat. Mater.}\ }\textbf {\bibinfo {volume}
  {16}},\ \bibinfo {pages} {977} (\bibinfo {year} {2017})}\BibitemShut
  {NoStop}%
\bibitem [{\citenamefont {Sheng}\ \emph {et~al.}(2017)\citenamefont {Sheng},
  \citenamefont {Sakuraba}, \citenamefont {Lau}, \citenamefont {Takahashi},
  \citenamefont {Mitani},\ and\ \citenamefont {Hayashi}}]{Sheng2017}%
  \BibitemOpen
  \bibfield  {author} {\bibinfo {author} {\bibfnamefont {P.}~\bibnamefont
  {Sheng}}, \bibinfo {author} {\bibfnamefont {Y.}~\bibnamefont {Sakuraba}},
  \bibinfo {author} {\bibfnamefont {Y.-C.}\ \bibnamefont {Lau}}, \bibinfo
  {author} {\bibfnamefont {S.}~\bibnamefont {Takahashi}}, \bibinfo {author}
  {\bibfnamefont {S.}~\bibnamefont {Mitani}},\ and\ \bibinfo {author}
  {\bibfnamefont {M.}~\bibnamefont {Hayashi}},\ }\href
  {https://doi.org/10.1126/sciadv.1701503} {\bibfield  {journal} {\bibinfo
  {journal} {Sci. Adv.}\ }\textbf {\bibinfo {volume} {3}},\ \bibinfo {pages}
  {e1701503} (\bibinfo {year} {2017})}\BibitemShut {NoStop}%
\bibitem [{\citenamefont {Bose}\ \emph
  {et~al.}(2018{\natexlab{a}})\citenamefont {Bose}, \citenamefont {Bhuktare},
  \citenamefont {Singh}, \citenamefont {Dutta}, \citenamefont {Achanta},\ and\
  \citenamefont {Tulapurkar}}]{Bose2018}%
  \BibitemOpen
  \bibfield  {author} {\bibinfo {author} {\bibfnamefont {A.}~\bibnamefont
  {Bose}}, \bibinfo {author} {\bibfnamefont {S.}~\bibnamefont {Bhuktare}},
  \bibinfo {author} {\bibfnamefont {H.}~\bibnamefont {Singh}}, \bibinfo
  {author} {\bibfnamefont {S.}~\bibnamefont {Dutta}}, \bibinfo {author}
  {\bibfnamefont {V.~G.}\ \bibnamefont {Achanta}},\ and\ \bibinfo {author}
  {\bibfnamefont {A.~A.}\ \bibnamefont {Tulapurkar}},\ }\href
  {https://doi.org/10.1063/1.5021731} {\bibfield  {journal} {\bibinfo
  {journal} {Appl. Phys. Lett.}\ }\textbf {\bibinfo {volume} {112}},\ \bibinfo
  {pages} {162401} (\bibinfo {year} {2018}{\natexlab{a}})}\BibitemShut
  {NoStop}%
\bibitem [{\citenamefont {{\v{Z}}elezn{\'{y}}}\ \emph
  {et~al.}(2017)\citenamefont {{\v{Z}}elezn{\'{y}}}, \citenamefont {Gao},
  \citenamefont {Manchon}, \citenamefont {Freimuth}, \citenamefont {Mokrousov},
  \citenamefont {Zemen}, \citenamefont {Ma{\v{s}}ek}, \citenamefont {Sinova},\
  and\ \citenamefont {Jungwirth}}]{Zelezny2017}%
  \BibitemOpen
  \bibfield  {author} {\bibinfo {author} {\bibfnamefont {J.}~\bibnamefont
  {{\v{Z}}elezn{\'{y}}}}, \bibinfo {author} {\bibfnamefont {H.}~\bibnamefont
  {Gao}}, \bibinfo {author} {\bibfnamefont {A.}~\bibnamefont {Manchon}},
  \bibinfo {author} {\bibfnamefont {F.}~\bibnamefont {Freimuth}}, \bibinfo
  {author} {\bibfnamefont {Y.}~\bibnamefont {Mokrousov}}, \bibinfo {author}
  {\bibfnamefont {J.}~\bibnamefont {Zemen}}, \bibinfo {author} {\bibfnamefont
  {J.}~\bibnamefont {Ma{\v{s}}ek}}, \bibinfo {author} {\bibfnamefont
  {J.}~\bibnamefont {Sinova}},\ and\ \bibinfo {author} {\bibfnamefont
  {T.}~\bibnamefont {Jungwirth}},\ }\href
  {https://doi.org/10.1103/PhysRevB.95.014403} {\bibfield  {journal} {\bibinfo
  {journal} {Phys. Rev. B}\ }\textbf {\bibinfo {volume} {95}},\ \bibinfo
  {pages} {14403} (\bibinfo {year} {2017})}\BibitemShut {NoStop}%
\bibitem [{\citenamefont {Dejene}\ \emph {et~al.}(2012)\citenamefont {Dejene},
  \citenamefont {Flipse},\ and\ \citenamefont {van Wees}}]{Dejene2012}%
  \BibitemOpen
  \bibfield  {author} {\bibinfo {author} {\bibfnamefont {F.~K.}\ \bibnamefont
  {Dejene}}, \bibinfo {author} {\bibfnamefont {J.}~\bibnamefont {Flipse}},\
  and\ \bibinfo {author} {\bibfnamefont {B.~J.}\ \bibnamefont {van Wees}},\
  }\href {https://doi.org/10.1103/PhysRevB.86.024436} {\bibfield  {journal}
  {\bibinfo  {journal} {Phys. Rev. B}\ }\textbf {\bibinfo {volume} {86}},\
  \bibinfo {pages} {024436} (\bibinfo {year} {2012})}\BibitemShut {NoStop}%
\bibitem [{\citenamefont {Blaha}\ \emph {et~al.}(2018)\citenamefont {Blaha},
  \citenamefont {Schwarz}, \citenamefont {Madsen}, \citenamefont {Kvasnicka},
  \citenamefont {Luitz}, \citenamefont {Laskowski}, \citenamefont {Tran},\ and\
  \citenamefont {Marks}}]{Blaha2018}%
  \BibitemOpen
  \bibfield  {author} {\bibinfo {author} {\bibfnamefont {P.}~\bibnamefont
  {Blaha}}, \bibinfo {author} {\bibfnamefont {K.}~\bibnamefont {Schwarz}},
  \bibinfo {author} {\bibfnamefont {G.~K.~H.}\ \bibnamefont {Madsen}}, \bibinfo
  {author} {\bibfnamefont {D.}~\bibnamefont {Kvasnicka}}, \bibinfo {author}
  {\bibfnamefont {J.}~\bibnamefont {Luitz}}, \bibinfo {author} {\bibfnamefont
  {R.}~\bibnamefont {Laskowski}}, \bibinfo {author} {\bibfnamefont
  {F.}~\bibnamefont {Tran}},\ and\ \bibinfo {author} {\bibfnamefont {L.~D.}\
  \bibnamefont {Marks}},\ }\href@noop {} {\emph {\bibinfo {title} {{WIEN2k, An
  Augmented Plane Wave + Local Orbitals Program for Calculating Crystal
  Properties (Karlheinz Schwarz, Techn. Universit{\"{a}}t Wien, Austria)}}}}\
  (\bibinfo {year} {2018})\BibitemShut {NoStop}%
\bibitem [{\citenamefont {Kubo}(1957)}]{Kubo1957}%
  \BibitemOpen
  \bibfield  {author} {\bibinfo {author} {\bibfnamefont {R.}~\bibnamefont
  {Kubo}},\ }\href@noop {} {\bibfield  {journal} {\bibinfo  {journal} {J. Phys.
  Soc. Jpn.}\ }\textbf {\bibinfo {volume} {12}},\ \bibinfo {pages} {570}
  (\bibinfo {year} {1957})}\BibitemShut {NoStop}%
\bibitem [{\citenamefont {Cutler}\ and\ \citenamefont
  {Mott}(1969)}]{cutlerObservationAndersonLocalization1969}%
  \BibitemOpen
  \bibfield  {author} {\bibinfo {author} {\bibfnamefont {M.}~\bibnamefont
  {Cutler}}\ and\ \bibinfo {author} {\bibfnamefont {N.~F.}\ \bibnamefont
  {Mott}},\ }\href@noop {} {\bibfield  {journal} {\bibinfo  {journal} {Phys.
  Rev.}\ }\textbf {\bibinfo {volume} {181}},\ \bibinfo {pages} {1336} (\bibinfo
  {year} {1969})}\BibitemShut {NoStop}%
\bibitem [{\citenamefont {Guo}\ \emph {et~al.}(2008)\citenamefont {Guo},
  \citenamefont {Murakami}, \citenamefont {Chen},\ and\ \citenamefont
  {Nagaosa}}]{Guo2008}%
  \BibitemOpen
  \bibfield  {author} {\bibinfo {author} {\bibfnamefont {G.~Y.}\ \bibnamefont
  {Guo}}, \bibinfo {author} {\bibfnamefont {S.}~\bibnamefont {Murakami}},
  \bibinfo {author} {\bibfnamefont {T.-W.}\ \bibnamefont {Chen}},\ and\
  \bibinfo {author} {\bibfnamefont {N.}~\bibnamefont {Nagaosa}},\ }\href
  {https://doi.org/10.1103/PhysRevLett.100.096401} {\bibfield  {journal}
  {\bibinfo  {journal} {Phys. Rev. Lett.}\ }\textbf {\bibinfo {volume} {100}},\
  \bibinfo {pages} {096401} (\bibinfo {year} {2008})}\BibitemShut {NoStop}%
\bibitem [{\citenamefont {Das}\ \emph {et~al.}(2017)\citenamefont {Das},
  \citenamefont {Schoemaker}, \citenamefont {van Wees},\ and\ \citenamefont
  {Vera-Marun}}]{Das2017}%
  \BibitemOpen
  \bibfield  {author} {\bibinfo {author} {\bibfnamefont {K.~S.}\ \bibnamefont
  {Das}}, \bibinfo {author} {\bibfnamefont {W.~Y.}\ \bibnamefont {Schoemaker}},
  \bibinfo {author} {\bibfnamefont {B.~J.}\ \bibnamefont {van Wees}},\ and\
  \bibinfo {author} {\bibfnamefont {I.~J.}\ \bibnamefont {Vera-Marun}},\ }\href
  {https://doi.org/10.1103/PhysRevB.96.220408} {\bibfield  {journal} {\bibinfo
  {journal} {Phys. Rev. B}\ }\textbf {\bibinfo {volume} {96}},\ \bibinfo
  {pages} {220408} (\bibinfo {year} {2017})}\BibitemShut {NoStop}%
\bibitem [{\citenamefont {Bose}\ \emph
  {et~al.}(2018{\natexlab{b}})\citenamefont {Bose}, \citenamefont {Lam},
  \citenamefont {Bhuktare}, \citenamefont {Dutta}, \citenamefont {Singh},
  \citenamefont {Jibiki}, \citenamefont {Goto}, \citenamefont {Miwa},\ and\
  \citenamefont {Tulapurkar}}]{Bose2018b}%
  \BibitemOpen
  \bibfield  {author} {\bibinfo {author} {\bibfnamefont {A.}~\bibnamefont
  {Bose}}, \bibinfo {author} {\bibfnamefont {D.~D.}\ \bibnamefont {Lam}},
  \bibinfo {author} {\bibfnamefont {S.}~\bibnamefont {Bhuktare}}, \bibinfo
  {author} {\bibfnamefont {S.}~\bibnamefont {Dutta}}, \bibinfo {author}
  {\bibfnamefont {H.}~\bibnamefont {Singh}}, \bibinfo {author} {\bibfnamefont
  {Y.}~\bibnamefont {Jibiki}}, \bibinfo {author} {\bibfnamefont
  {M.}~\bibnamefont {Goto}}, \bibinfo {author} {\bibfnamefont {S.}~\bibnamefont
  {Miwa}},\ and\ \bibinfo {author} {\bibfnamefont {A.~A.}\ \bibnamefont
  {Tulapurkar}},\ }\href {https://doi.org/10.1103/PhysRevApplied.9.064026}
  {\bibfield  {journal} {\bibinfo  {journal} {Phys. Rev. Applied}\ }\textbf
  {\bibinfo {volume} {9}},\ \bibinfo {pages} {064026} (\bibinfo {year}
  {2018}{\natexlab{b}})}\BibitemShut {NoStop}%
\bibitem [{\citenamefont {Gibbons}\ \emph {et~al.}(2018)\citenamefont
  {Gibbons}, \citenamefont {MacNeill}, \citenamefont {Buhrman},\ and\
  \citenamefont {Ralph}}]{Gibbons2018}%
  \BibitemOpen
  \bibfield  {author} {\bibinfo {author} {\bibfnamefont {J.~D.}\ \bibnamefont
  {Gibbons}}, \bibinfo {author} {\bibfnamefont {D.}~\bibnamefont {MacNeill}},
  \bibinfo {author} {\bibfnamefont {R.~A.}\ \bibnamefont {Buhrman}},\ and\
  \bibinfo {author} {\bibfnamefont {D.~C.}\ \bibnamefont {Ralph}},\ }\href
  {https://doi.org/10.1103/PhysRevApplied.9.064033} {\bibfield  {journal}
  {\bibinfo  {journal} {Phys. Rev. Applied}\ }\textbf {\bibinfo {volume} {9}},\
  \bibinfo {pages} {064033} (\bibinfo {year} {2018})}\BibitemShut {NoStop}%
\bibitem [{\citenamefont {Iihama}\ \emph {et~al.}(2018)\citenamefont {Iihama},
  \citenamefont {Taniguchi}, \citenamefont {Yakushiji}, \citenamefont
  {Fukushima}, \citenamefont {Shiota}, \citenamefont {Tsunegi}, \citenamefont
  {Hiramatsu}, \citenamefont {Yuasa}, \citenamefont {Suzuki},\ and\
  \citenamefont {Kubota}}]{Iihama2018}%
  \BibitemOpen
  \bibfield  {author} {\bibinfo {author} {\bibfnamefont {S.}~\bibnamefont
  {Iihama}}, \bibinfo {author} {\bibfnamefont {T.}~\bibnamefont {Taniguchi}},
  \bibinfo {author} {\bibfnamefont {K.}~\bibnamefont {Yakushiji}}, \bibinfo
  {author} {\bibfnamefont {A.}~\bibnamefont {Fukushima}}, \bibinfo {author}
  {\bibfnamefont {Y.}~\bibnamefont {Shiota}}, \bibinfo {author} {\bibfnamefont
  {S.}~\bibnamefont {Tsunegi}}, \bibinfo {author} {\bibfnamefont
  {R.}~\bibnamefont {Hiramatsu}}, \bibinfo {author} {\bibfnamefont
  {S.}~\bibnamefont {Yuasa}}, \bibinfo {author} {\bibfnamefont
  {Y.}~\bibnamefont {Suzuki}},\ and\ \bibinfo {author} {\bibfnamefont
  {H.}~\bibnamefont {Kubota}},\ }\href
  {https://doi.org/10.1038/s41928-018-0026-z} {\bibfield  {journal} {\bibinfo
  {journal} {Nat. Electr.}\ }\textbf {\bibinfo {volume} {1}},\ \bibinfo {pages}
  {120} (\bibinfo {year} {2018})}\BibitemShut {NoStop}%
\bibitem [{\citenamefont {Seki}\ \emph {et~al.}(2019)\citenamefont {Seki},
  \citenamefont {Iihama}, \citenamefont {Taniguchi},\ and\ \citenamefont
  {Takanashi}}]{Seki2019}%
  \BibitemOpen
  \bibfield  {author} {\bibinfo {author} {\bibfnamefont {T.}~\bibnamefont
  {Seki}}, \bibinfo {author} {\bibfnamefont {S.}~\bibnamefont {Iihama}},
  \bibinfo {author} {\bibfnamefont {T.}~\bibnamefont {Taniguchi}},\ and\
  \bibinfo {author} {\bibfnamefont {K.}~\bibnamefont {Takanashi}},\ }\href
  {https://doi.org/10.1103/PhysRevB.100.144427} {\bibfield  {journal} {\bibinfo
   {journal} {Phys. Rev. B}\ }\textbf {\bibinfo {volume} {100}},\ \bibinfo
  {pages} {144427} (\bibinfo {year} {2019})}\BibitemShut {NoStop}%
\bibitem [{\citenamefont {Tanaka}\ \emph
  {et~al.}(2008{\natexlab{b}})\citenamefont {Tanaka}, \citenamefont {Kontani},
  \citenamefont {Naito}, \citenamefont {Naito}, \citenamefont {Hirashima},
  \citenamefont {Yamada},\ and\ \citenamefont {Inoue}}]{Tanaka2008}%
  \BibitemOpen
  \bibfield  {author} {\bibinfo {author} {\bibfnamefont {T.}~\bibnamefont
  {Tanaka}}, \bibinfo {author} {\bibfnamefont {H.}~\bibnamefont {Kontani}},
  \bibinfo {author} {\bibfnamefont {M.}~\bibnamefont {Naito}}, \bibinfo
  {author} {\bibfnamefont {T.}~\bibnamefont {Naito}}, \bibinfo {author}
  {\bibfnamefont {D.~S.}\ \bibnamefont {Hirashima}}, \bibinfo {author}
  {\bibfnamefont {K.}~\bibnamefont {Yamada}},\ and\ \bibinfo {author}
  {\bibfnamefont {J.}~\bibnamefont {Inoue}},\ }\href
  {https://doi.org/10.1103/PhysRevB.77.165117} {\bibfield  {journal} {\bibinfo
  {journal} {Phys. Rev. B}\ }\textbf {\bibinfo {volume} {77}},\ \bibinfo
  {pages} {165117} (\bibinfo {year} {2008}{\natexlab{b}})}\BibitemShut
  {NoStop}%
\bibitem [{\citenamefont {Qu}\ \emph {et~al.}(2020)\citenamefont {Qu},
  \citenamefont {Nakamura},\ and\ \citenamefont
  {Hayashi}}]{quMagnetizationDirectionDependent2020}%
  \BibitemOpen
  \bibfield  {author} {\bibinfo {author} {\bibfnamefont {G.}~\bibnamefont
  {Qu}}, \bibinfo {author} {\bibfnamefont {K.}~\bibnamefont {Nakamura}},\ and\
  \bibinfo {author} {\bibfnamefont {M.}~\bibnamefont {Hayashi}},\ }\href@noop
  {} {\bibfield  {journal} {\bibinfo  {journal} {Phys. Rev. B}\ }\textbf
  {\bibinfo {volume} {102}},\ \bibinfo {pages} {144440} (\bibinfo {year}
  {2020})}\BibitemShut {NoStop}%
\bibitem [{\citenamefont {Miura}\ and\ \citenamefont
  {Masuda}(2021)}]{Miura2021}%
  \BibitemOpen
  \bibfield  {author} {\bibinfo {author} {\bibfnamefont {Y.}~\bibnamefont
  {Miura}}\ and\ \bibinfo {author} {\bibfnamefont {K.}~\bibnamefont {Masuda}},\
  }\href {https://doi.org/10.1103/PhysRevMaterials.5.L101402} {\bibfield
  {journal} {\bibinfo  {journal} {Phys. Rev. Materials}\ }\textbf {\bibinfo
  {volume} {5}},\ \bibinfo {pages} {L101402} (\bibinfo {year}
  {2021})}\BibitemShut {NoStop}%
\bibitem [{\citenamefont {Oppeneer}(2001)}]{Oppeneer2001}%
  \BibitemOpen
  \bibfield  {author} {\bibinfo {author} {\bibfnamefont {P.~M.}\ \bibnamefont
  {Oppeneer}},\ }in\ \href@noop {} {\emph {\bibinfo {booktitle} {Handbook of
  Magnetic Materials}}},\ Vol.~\bibinfo {volume} {13},\ \bibinfo {editor}
  {edited by\ \bibinfo {editor} {\bibfnamefont {K.~H.~J.}\ \bibnamefont
  {Buschow}}}\ (\bibinfo  {publisher} {Elsevier, Amsterdam},\ \bibinfo {year}
  {2001})\ pp.\ \bibinfo {pages} {229 -- 422}\BibitemShut {NoStop}%
\bibitem [{\citenamefont {Omori}\ \emph {et~al.}(2019)\citenamefont {Omori},
  \citenamefont {Sagasta}, \citenamefont {Niimi}, \citenamefont {Gradhand},
  \citenamefont {Hueso}, \citenamefont {Casanova},\ and\ \citenamefont
  {Otani}}]{Omori2019}%
  \BibitemOpen
  \bibfield  {author} {\bibinfo {author} {\bibfnamefont {Y.}~\bibnamefont
  {Omori}}, \bibinfo {author} {\bibfnamefont {E.}~\bibnamefont {Sagasta}},
  \bibinfo {author} {\bibfnamefont {Y.}~\bibnamefont {Niimi}}, \bibinfo
  {author} {\bibfnamefont {M.}~\bibnamefont {Gradhand}}, \bibinfo {author}
  {\bibfnamefont {L.~E.}\ \bibnamefont {Hueso}}, \bibinfo {author}
  {\bibfnamefont {F.}~\bibnamefont {Casanova}},\ and\ \bibinfo {author}
  {\bibfnamefont {Y.}~\bibnamefont {Otani}},\ }\href
  {https://doi.org/10.1103/PhysRevB.99.014403} {\bibfield  {journal} {\bibinfo
  {journal} {Phys. Rev. B}\ }\textbf {\bibinfo {volume} {99}},\ \bibinfo
  {pages} {014403} (\bibinfo {year} {2019})}\BibitemShut {NoStop}%
\bibitem [{\citenamefont {Nagaosa}\ \emph {et~al.}(2010)\citenamefont
  {Nagaosa}, \citenamefont {Sinova}, \citenamefont {Onoda}, \citenamefont
  {MacDonald},\ and\ \citenamefont {Ong}}]{Nagaosa2010}%
  \BibitemOpen
  \bibfield  {author} {\bibinfo {author} {\bibfnamefont {N.}~\bibnamefont
  {Nagaosa}}, \bibinfo {author} {\bibfnamefont {J.}~\bibnamefont {Sinova}},
  \bibinfo {author} {\bibfnamefont {S.}~\bibnamefont {Onoda}}, \bibinfo
  {author} {\bibfnamefont {A.~H.}\ \bibnamefont {MacDonald}},\ and\ \bibinfo
  {author} {\bibfnamefont {N.~P.}\ \bibnamefont {Ong}},\ }\href
  {https://doi.org/10.1103/RevModPhys.82.1539} {\bibfield  {journal} {\bibinfo
  {journal} {Rev. Mod. Phys.}\ }\textbf {\bibinfo {volume} {82}},\ \bibinfo
  {pages} {1539} (\bibinfo {year} {2010})}\BibitemShut {NoStop}%
\end{thebibliography}%

\end{document}